\documentclass[twocolumn,superscriptaddress,nofootinbib]{revtex4-2}
\pdfoutput=1

\usepackage[utf8]{inputenc}
\usepackage{comment}
\usepackage[english]{babel}
\usepackage{amsfonts, amsmath, amssymb, amsthm}
\usepackage[colorlinks=true,breaklinks=true,linkcolor=black,citecolor=green,urlcolor=magenta]{hyperref}
\usepackage{graphicx}
\usepackage[export]{adjustbox}

\theoremstyle{remark}

\usepackage{xcolor}
\usepackage[capitalise]{cleveref}
\usepackage{algorithmicx}
\usepackage{algorithm}
\usepackage[noend]{algpseudocode}
\usepackage[normalem]{ulem}
\usepackage{booktabs}

\algnewcommand\algorithmicinput{\textbf{Input:}}
\algnewcommand\Input{\item[\algorithmicinput]}

\algnewcommand\algorithmicoutput{\textbf{Output:}}
\algnewcommand\Output{\item[\algorithmicoutput]}

\usepackage{bm}
\usepackage{braket}

\newcommand{\device}[1]{$\mathsf{ibm\_#1}$}

\newcommand{\nocontentsline}[3]{}
\newcommand{\tocless}[2]{\bgroup\let\addcontentsline=\nocontentsline#1{#2}\egroup}

\begin{document}

\title{Practical and efficient quantum circuit synthesis and transpiling with Reinforcement Learning}
\author{David Kremer}
\email{david.kremer@ibm.com}
\affiliation{IBM Quantum, IBM T.J. Watson Research Center, Yorktown Heights, NY 10598}
\author{Victor Villar}
\affiliation{IBM Quantum, IBM T.J. Watson Research Center, Yorktown Heights, NY 10598}
\author{Hanhee Paik}
\affiliation{IBM Quantum, IBM T.J. Watson Research Center, Yorktown Heights, NY 10598}
\author{Ivan Duran}
\affiliation{IBM Quantum, IBM T.J. Watson Research Center, Yorktown Heights, NY 10598}
\author{Ismael Faro}
\affiliation{IBM Quantum, IBM T.J. Watson Research Center, Yorktown Heights, NY 10598}
\author{Juan Cruz-Benito}
\email{juan.cruz.benito@ibm.com}
\affiliation{IBM Quantum, IBM T.J. Watson Research Center, Yorktown Heights, NY 10598}


\maketitle

\textbf{
This paper demonstrates the integration of Reinforcement Learning (RL) into quantum transpiling workflows, significantly enhancing the synthesis and routing of quantum circuits. By employing RL, we achieve near-optimal synthesis of Linear Function, Clifford, and Permutation circuits, up to 9, 11 and 65 qubits respectively, while being compatible with native device instruction sets and connectivity constraints, and orders of magnitude faster than optimization methods such as SAT solvers. We also achieve significant reductions in two-qubit gate depth and count for circuit routing up to 133 qubits with respect to other routing heuristics such as SABRE. We find the method to be efficient enough to be useful in practice in typical quantum transpiling pipelines. Our results set the stage for further AI-powered enhancements of quantum computing workflows.
}

\tocless\section{Introduction}\label{section:introduction}

The field of quantum computing has grown into a community that includes domain experts in a wide range of various disciplines such as chemistry and Artificial Intelligence (AI) from industry, government and academia.  With the rapidly developing technologies, quantum computing is now accessible by users via cloud where users are exploring how to utilize quantum computing as a tool for their goals.  In 2023, IBM has demonstrated that with 127-qubit Eagle processor and error mitigation techniques, we can execute quantum circuits outside the reach of classical exact computing methods – the beginning of quantum utility \cite{kim2023evidence}. Now quantum computing is evolving to expand its computational power by integrating with classical computing resources. Exploiting classical resources will allow us to extend the computational capacity that the current quantum hardware can offer and increase the workloads that are required for practical quantum computing. The future of computing would be a computing system where the orchestration between classical computing resources such as CPUs and GPUs together with multiple quantum processors (QPUs) \cite{alexeev2023quantum,robledo2024chemistry}, forms a composite architecture that allows for scaling quantum computation to sizes amenable to industry scale problems on the order of tens of thousands of qubits \cite{bravyi2022future}.

The integration of AI tools is considered a critical path to advance in several scientific fields \cite{jumper2021highly, fawzi2022discovering, schwaller2019molecular,das2021accelerated, wang2023scientific,merchant2023scaling,trinh2024solving}. In particular, we observe AI has the potential to help leverage the full potential of quantum computing \cite{dunjko2018machine}. As the sizes of quantum circuits input to quantum computing systems continue to grow in the era of quantum utility, there is a need for smarter and more effective methods to enhance the quantum computation workflow. The integration of classical computing resources, such as GPUs and CPUs, with QPUs will pave the way for new frontiers in computing, and the collaboration between AI and quantum computing will be critical in realizing this potential. 

Quantum circuit transpilation and optimization are central components of quantum computing workflows. Similar to compilers in classical computing, transpilers map logical quantum circuits to the instructions present physical quantum devices, and allow quantum circuit developers to focus on the quantum algorithms rather than specific details of the hardware. High-quality transpilation of quantum circuits (minimizing the overhead introduced in this mapping) is an important goal for the field of quantum computing in general, but especially relevant for near-term quantum computing hardware, where even small improvements on the transpiled circuit sizes can lead to important reductions in the noise present in the results. 

Circuit synthesis is a useful task within transpiling workflows and consists on generating a quantum circuit, within a given set of gates, that implements a high-level description of a quantum operator. A typical workflow with circuit synthesis is to re-synthesize parts of a circuit to see if a ``better" circuit can be found and replace the original circuit parts if successful. 

The practical usefulness of such workflow relies on having access to synthesis methods that can produce optimal or near-optimal circuits, with a reasonable amount of computational resources. Circuit synthesis methods typically fall in three categories:
\begin{itemize}
    \item \textbf{Heuristic methods:} Heuristic methods are typically fast, but often produce sub-optimal circuits and work with a specific gate set such that changing a gate set requires modifying the algorithm or incurs a further translation cost.
    \item \textbf{Databases of optimal circuits:} Databases of optimal circuits can provide optimal circuits quickly, but take a long time to generate and require large data storage. Also fixed to specific gate sets.
    \item \textbf{Generic optimization:} produces optimal circuits tailored to a specific gate set, but at the cost of high running times.
\end{itemize}

Circuit routing is also a major task in typical transpiling workflows, and consists on inserting SWAP operations on a circuit to make two-qubit operations compatible with a given coupling map that restricts the pairs of qubits on which operations can be applied. Circuit routing faces similar challenges as circuit synthesis: there are fast heuristic algorithms but the resulting circuits are often far from optimal routing, especially in terms of circuit depth, and optimization methods have prohibitively high computational cost.

In the literature, numerous algorithms and heuristic methods have been proposed for synthesizing Clifford circuits for full connectivity, such as \cite{bravyi2021clifford,bravyi2021hadamard,aaronson2004improved}, databases of optimal Clifford circuits up to 6 qubits \cite{bravyi20226}, and algorithims for synthesis of Clifford circuits on linear connectivity such as \cite{maslov2018shorter}. However, these methods do not work for other types of restricted connectivity graphs, so further routing is needed to run on real devices with architectures such as heavy-hex, resulting in a significant SWAP overhead. There are also full optimization methods based on SAT solvers that offer complete flexibility in terms of connectivity but scale exponentially with the number of qubits and circuit sizes and are not practical for larger circuits \cite{schneider2023sat}. 
For Permutation circuits, efficient algorithms exist for synthesizing Permutation circuits with full connectivity \cite{alon1993routing} and linear connectivity \cite{kutin2007computation}. There are also heuristic methods that can implement permutations while fulfilling connectivity constraints on arbitrary graphs \cite{wagner2023improving} but as we show here, produce results that are far from optimal, especially for circuit depth. Again, for optimal or near optimal synthesis on arbitrary connectivity one has to resort to optimization methods such as \cite{chen2022recursive,berent2022towards}. 
For Linear Function circuits, efficient optimal methods exist to synthesize Linear Functions without connectivity restrictions \cite{markov2008optimal}, but the resulting circuits are far from optimal after routing to real device topologies. Efficient methods for synthesizing Linear Functions on linear connectivity also exist \cite{kutin2007computation}, but they are not optimal in terms of CNOT count and, once again, far from optimal if the circuit needs routing to other topologies. Again, full optimization methods based on SAT solvers that offer complete flexibility in terms of connectivity are available, but they scale exponentially and are not practical beyond 10-12 qubits \cite{chen2022recursive}. 
In the context of qubit routing, computationally efficient algorithms exist that can be used for larger systems (127+ qubits) \cite{li2019tackling} but they produce transpiled circuits that are far from optimal in terms of circuit depth, and they cannot directly leverage local circuit optimizations (e.g., cancellation of two consecutive CNOT gates). Full optimization methods based on SAT solvers can produce optimal or near-optimal circuits and can leverage gate optimizations, but they scale exponentially and are not feasible in a practical environment beyond 8-10 qubits \cite{nannicini2022optimal}. 

In terms of AI applied to quantum circuits, the literature offers several examples in the areas of synthesis  \cite{murakami2022autoqc,arrazola2019machine,bu2025physical}, optimization \cite{fosel2021quantum,ruiz2024quantum,lockwood2021optimizing,zen2024quantum}, mapping \cite{acampora2021deep,paler2023machine} or compilation \cite{moro2021quantum,arufe2022quantum,rasconi2019innovative,quetschlich2023predicting,he2021variational,quetschlich2023compiler}. This paper builds upon existing state-of-the-art procedures and focuses on applying AI techniques to quantum circuit transpilation and optimization. 

We introduce a generic Reinforcement Learning(RL)-based approach to synthesize various types of quantum circuits, including Clifford, Linear Function, and Permutation circuits. Our RL-based synthesis method generates near-optimal results and generates circuits directly compatible with the native instruction set of the device (including connectivity restrictions), so the resulting circuits can be directly executed in quantum devices without the need for additional transpilation, and can be as an optimization step in the final stages of a transpiling pipeline. By striking a balance between circuit optimality, adaptability, and computational cost, our approach makes circuit synthesis a valuable tool for practical transpiling workflows. We extend this method to circuit routing, yielding significant enhancements over existing heuristic techniques while incurring much lower computational costs than generic optimization. Our key finding is that our RL-based methods enhance the current heuristic algorithms used in libraries like Qiskit SDK \cite{Qiskit,javadi2024quantum} without significantly increasing time performance, while providing substantial time efficiency improvements compared to existing hard optimization methods, such as SAT solvers. These results pave the way for the integration of these methods into real quantum computing environments as essential components of transpilation and execution workflows.


\vspace{2em}
\tocless\section{Results}\label{section:Results}

\subsection{Circuit Synthesis with Reinforcement Learning}

Our method is based on applying Reinforcement Learning to circuit synthesis by framing circuit synthesis as a sequential decision process. 

\begin{figure*}[tp]
    \centering
    \includegraphics[width=1\linewidth]{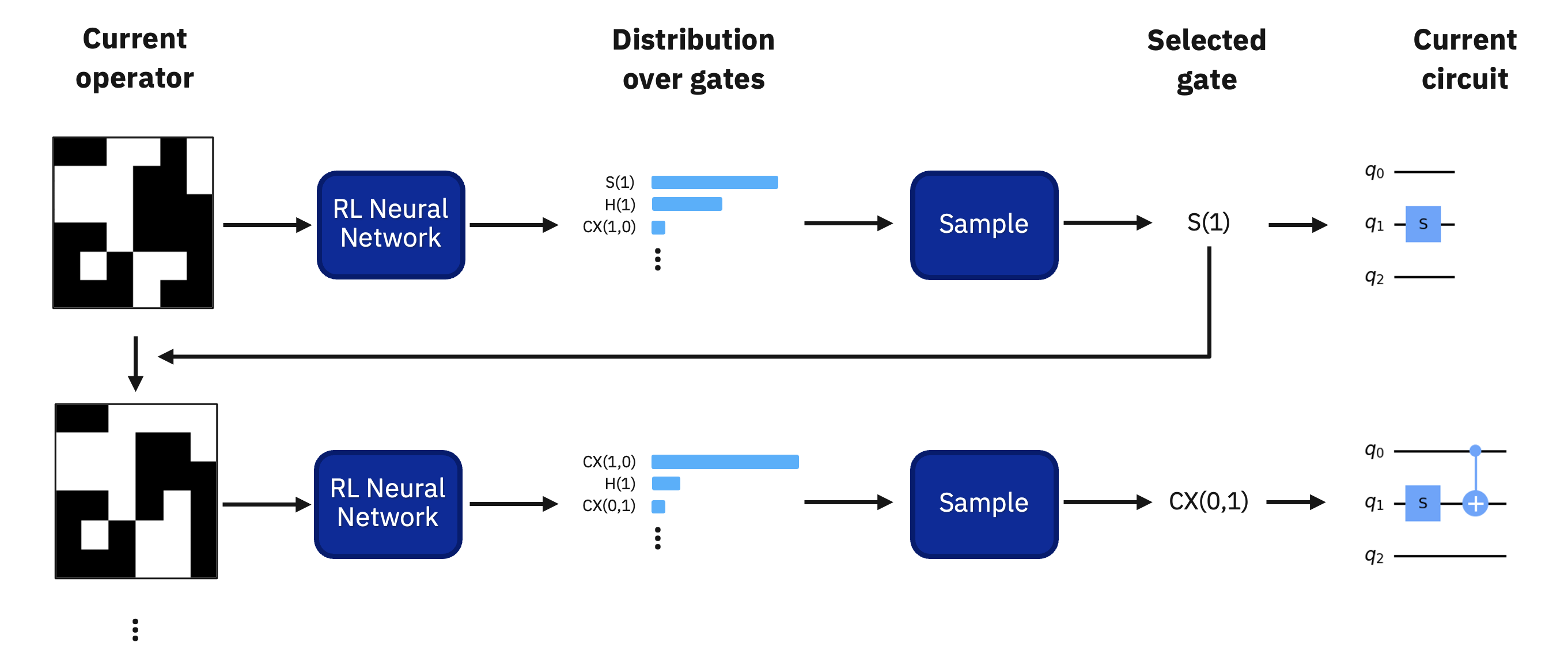}
    \caption{Diagram describing the RL-based circuit synthesis process. }
    \label{fig:enter-label}
\end{figure*}

Circuit synthesis can be described as a sequential decision process where one decides at each step, for an operator $O_t$, which of the possible operations of the gate set to apply. Once a specific gate has been chosen ($g_t$), the operator is evolved based on the gate to obtain the next operator $O_{t+1}$. Starting from the input operator $O_0$ (the operator we want to implement as a circuit) we repeat the decision process for a given number of steps $T$ until we reach the identity operator $O_T = I$. The circuit that implements $O_0$ can be then recovered by inverting the circuit given by the sequence of gates $g_{0..T}$. A depiction of this process is shown in Figure~\ref{fig:enter-label}.

At the core of the algorithm lays the RL agent, which is responsible for selecting the gate $g_t$ at each step for the given operator $O_t$. As is usual with other AI methods, our method requires an initial \emph{training} phase, where the RL agent learns to perform the circuit synthesis, and after that, it can be used for \emph{inference}, where the RL agent applies the knowledge it has gathered to synthesize actual circuits. Importantly, the training phase only needs to happen once for a given class of operators and a given gate set (including particular connectivity restrictions).

\subsubsection{Training}

For training the RL agent, we follow the standard RL procedure where we let the RL agent try to synthesize input operators and learn from the results, but with the addition of dynamically adjusting the difficulty of the input operators as the RL agent gets better (this is sometimes called ``curriculum learning'' in the RL literature \cite{narvekar2020curriculum}). 

The way the agent is provided feedback of its actions is through a reward function. At each decision at step $t$, a reward function $r_t$ is generated based on the immediate results of that action. In our method, the reward function has two components:
\begin{enumerate}
    \item A large positive reward when the decision results in reaching the identity operator (meaning that the synthesis has been completed), to encourage the agent to choose actions that lead to synthesizing the circuit.
    \item Small negative rewards (penalties) for the gates applied at that step, to encourage the agent to choose actions that lead to lower gate counts / circuit depths. Different gates could incur different penalties (typically, two-qubit gates are much more expensive than single-qubit gates). These penalties can also depend on more global properties of the circuit, such as if the gate applied resulted on an increase of the total circuit depth.
\end{enumerate}
The RL algorithm uses this reward at each step of the training to adjust the network weights with the aim of maximizing the cumulative reward function. 

The full procedure for training is the following:
\begin{enumerate}
    \item Start by initializing the network weights (with some standard neural network initialization method such as Xavier \cite{glorot2010understanding}) and setting the difficulty to $1$.
    \item At each training step (for a fixed number of steps):
    \begin{enumerate}
        \item Generate a batch of target operators of difficulty $d$.
        \item Run the inference process for the target operators, collecting the $O_t$, $g_t$ and $r_t$ generated at each step.
        \item Adjust the network weights following the chosen RL algorithm with the data collected.
        \item If the success rate (that is, the rate of target operators that were successfully synthesized) is above a given threshold, increase the difficulty $d$ by 1.
    \end{enumerate}
\end{enumerate}

We found that dynamically adjusting the difficulty of the target operators was crucial for the training to converge. To generate a target operator of a given difficulty $d$, we just generate a random circuit by sampling $d$ gates from the target gate set, and we compute the operator resulting from that circuit. This ensures that there is a solution that uses at most $d$ gates from the gate set.

Note that, as with typical RL methods, the RL agent does not learn from a ``labeled" dataset, that is, it doesn't learn from pairs of operators-circuits: it learns to synthesize the circuits only by trial and error and the guidance of the reward function, without any previous knowledge of circuit synthesis. This allows the RL method to potentially outperform existing methods, as it is not mimicking existing heuristics; instead, it is exploring the space of possible heuristics to find ones that yield high rewards. This also allows the method to scale beyond the reach of databases of optimal circuits, as these are not needed for training. 

\subsubsection{Inference}

Once an RL agent has been trained, we can synthesize a given operator $O_0$ by following the decision process described earlier and where the decisions on which gate $g_t$ is chosen at each step are taken by the RL agent. In our implementation, the RL agent consists of just a neural network that takes the given operator $O_t$ as input (in some numerical representation) and outputs a log-likelihood for each of the possible gates. 

Given these log-likelihoods, one can follow different strategies to generate the circuits, similarly to what \textit{decoders} do for Large Language Models \cite{shi2024thorough}:
\begin{itemize}
    \item \textbf{Greedy:} where, at each step, we just pick the gate with the highest likelihood. This will deterministically produce a circuit for the given input operator.
    \item \textbf{Sampling:} at each step, we sample from the probability distribution generated by the RL agent. This allows us to sample different circuits that implement the given input operator (of potentially different quality).
    \item \textbf{Searching:} we search the decision tree by selecting the gates with "top k" likelihoods, or the gates with likelihoods above a given threshold (also known as "top p" for LLMs). As with sampling, this will also produce different circuits for the given input operator, but may be more efficient than sampling if we aim to sample a large number of circuits.
\end{itemize}

Interestingly, we have found that just the operator $O_t$ is enough input for the network to produce gate choices at each step that lead to high-quality circuits, that is, the network does not need to know anything about the operators from previous steps or the gates chosen so far. 

Note that nothing guarantees that the algorithm is successful in implementing the given operator in a finite number of steps. In fact, if we select the gates deterministically (Greedy method), it is easy to see that the iteration might enter a closed loop where the same operators are visited in a sequence. Interestingly, however, in our experiments the fully-trained agents achieve an 100\% success rate for a fixed and reasonable step limit $T$.

In the following subsections we show how we apply this method Clifford and permutation circuit synthesis, and to circuit routing.

\subsubsection{Clifford Circuit Synthesis with Reinforcement Learning}

Here we apply the methods previously described to Clifford circuit synthesis with constrained two-qubit gate connectivity, and optimizing for two-qubit gate count and depth.

As described in \cite{aaronson2004improved}, a Clifford circuit of $N$ qubits can be represented as a $2N$ by $2N$ boolean matrix plus a $2N$ boolean phase vector that can be updated in $O(N)$ time when applying elementary Clifford gates (H, S, CX) (often referred as ``the Clifford tableau"). For simplicity, we will just focus on the matrix and disregard the phase vector, since the phase can always be corrected after the synthesis by adding a limited number of single-qubit gates. 

As an example, we show a model trained for 7 qubits with ``H" connectivity (all the topology names used in this paper are described in supplementary section \ref{section:SupplementalInfoCouplingMaps} ). A 7-qubit Clifford can be represented as a $14$ by $14$ boolean matrix (disregarding the phase). Here, we reshape the matrix as four $7$ by $7$ matrices that we stack on a third dimension, forming a $7$ by $7$ by $4$ tensor, and use this as input to our network.

The output is given by all the possible Clifford gates that can be applied within the ``H" connectivity. Here we use H, S and CNOT gates as the basis gates. Taking into account the connectivity limitations, and allowing CNOTs in both directions, we have a total of $7 + 7 + 2 * 6 = 26$ possible actions. For this model, we define a network architecture consisting of a first convolutional layer followed by two fully connected layers. The model is around $1.3$ million parameters (with a disk size of $\sim$5 megabytes).

Starting from difficulty $1$, we train the model for $1024$ difficulty levels, where we generate the target Cliffords at each difficulty $d$ by sampling $d$ gates from the gate set. Once the maximum difficulty is reached, we uniformly sample from the space of 7 qubit Cliffords using the method described by \cite{bravyi2021hadamard}.

\begin{figure*}[tp]
    \centering
    \includegraphics[width=1\linewidth]{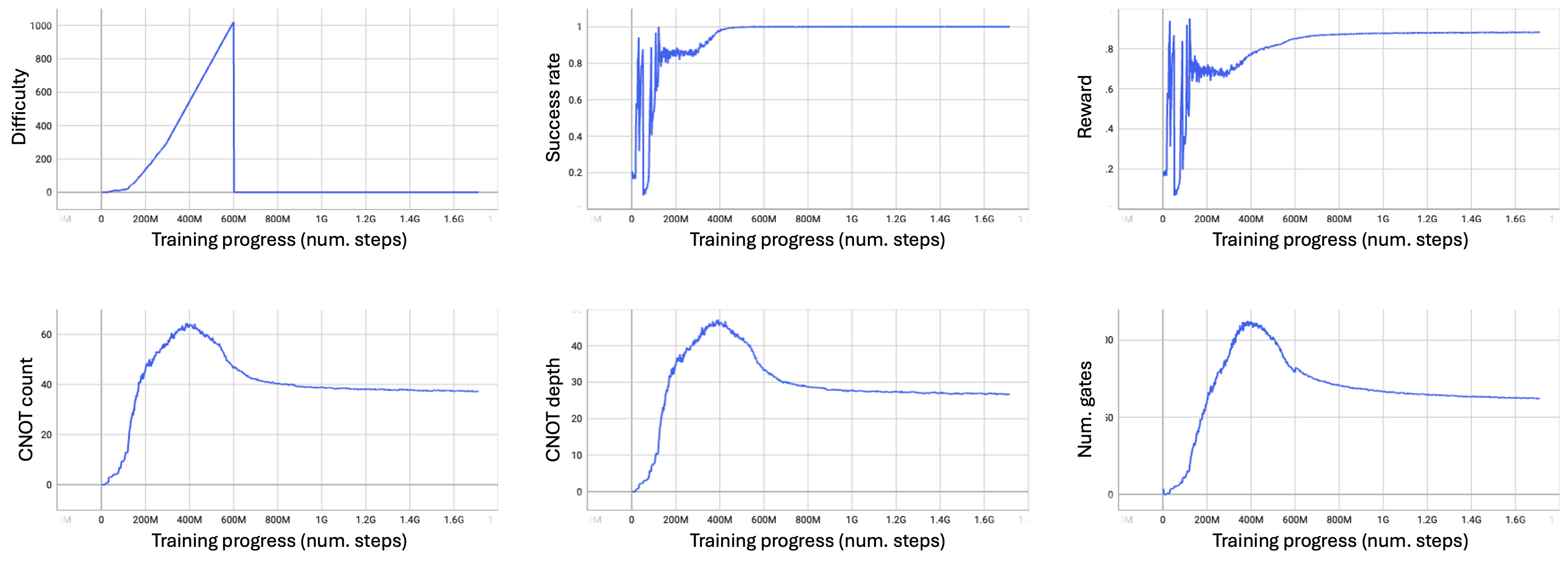}
    \caption{Training progress for Clifford synthesis on 7 qubits with ``H" connectivity. The horizontal axis shows the progress of the training in terms of total number of steps taken (number of Cliffords ``seen" by the model). The vertical axes on the different graphs represent how different quantities evolve through the training.}
    \label{fig:Training_process_7qH}
\end{figure*}

The progress of the training is shown in Figure \ref{fig:Training_process_7qH}. The training shows a pattern that we observe in most of the synthesis models, and follows three phases:
\begin{enumerate}
    \item An initial increase in success rate at difficulty 1, until the threshold, where the model learns to recognize the basic few-gate Cliffords.
    \item A plateau where the difficulty progressively increases as the model learns to synthesize harder Cliffords.
    \item A final phase where the model reaches $100\%$ success rate and progressively improves the quality of the synthesized circuits.
\end{enumerate}

Note that the total number of Cliffords seen by the model (shown in the ``x" axis of the graphs) is several orders of magnitude lower than the total number of 7-qubit Cliffords, and this is without discarding duplicates during the training. For reference, \cite{bravyi20226} shows an optimal database of Clifford circuits in 6 qubits took around two months to compute and weighted $2.1$ TB. Despite this, the model learns to virtually synthesize $100\%$ of Clifford circuits and, as we show on the Benchmark section, to do so with near optimal CNOT count and depth.

\subsubsection{Circuit Synthesis Benchmarks}

We benchmark our Clifford and permutation synthesis models against a SAT solver, and against heuristic algorithms for reference.

\begin{figure*}[tbp]
    \centering
    \begin{minipage}{0.48\textwidth}
        \centering
        \includegraphics[width=1.0\linewidth]{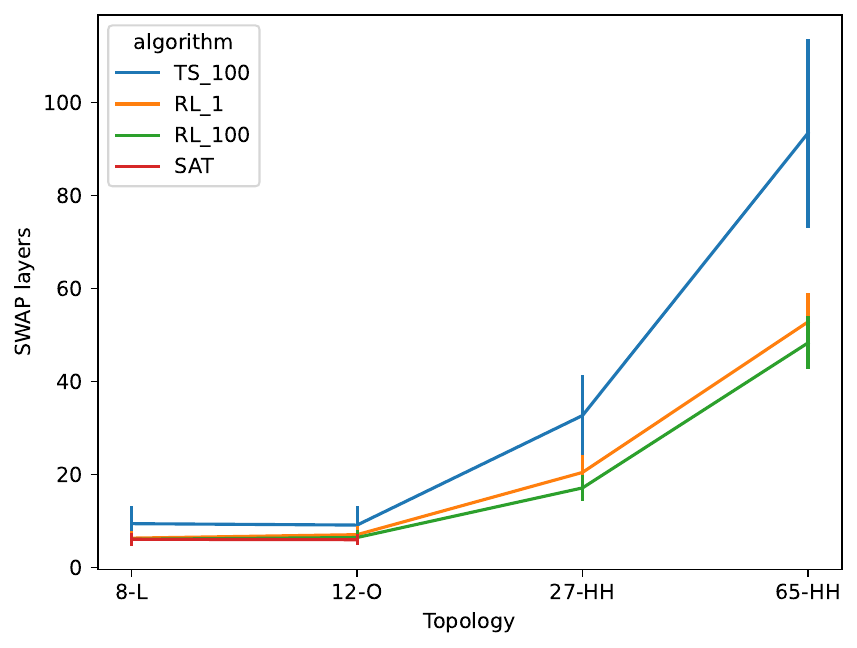}
        \caption{Number of SWAP layers obtained from synthesizing random permutations for 4 different topologies on 8, 12, 27 and 65 qubits, with an heuristic algorithm (TokenSwapper with 100 trials), a SAT solver, and our RL algorithm with 1 and 100 runs.}
        \label{fig:perms_vs_SAT}
    \end{minipage}\hfill
    \begin{minipage}{0.48\textwidth}
        \centering
        \includegraphics[width=1.0\linewidth]{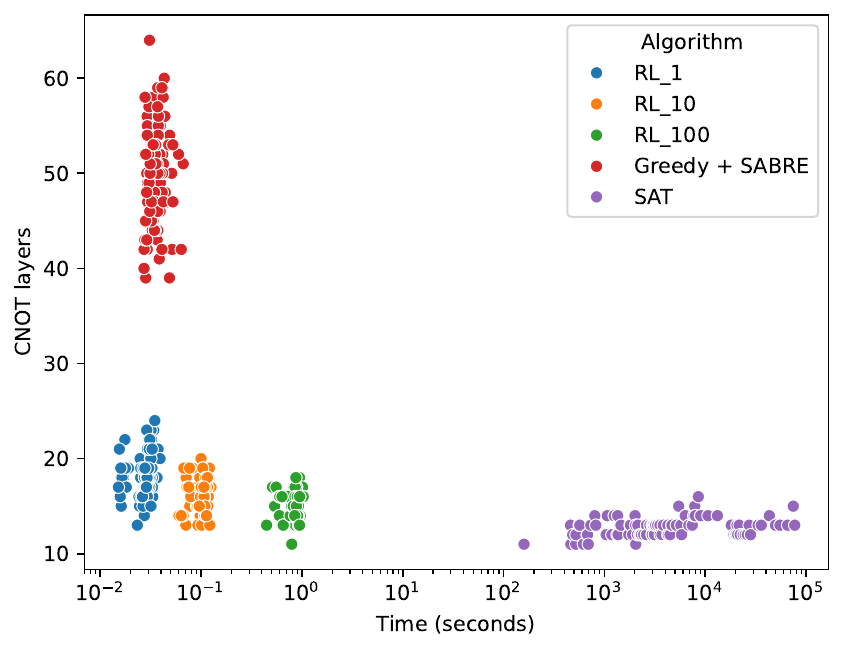}
        \caption{Number of CNOT layers obtained from synthesizing 100 random Cliffords for the 6-Y topology, with an heuristic algorithm (Greedy with further mapping using SABRE), a SAT solver, and our RL algorithm with 1, 10 and 100 runs, against the time taken by each algorithm for each of the Cliffords.}
        \label{fig:clifford_6qY_vs_SAT}
    \end{minipage}
\end{figure*}

For permutation synthesis, as shown in Figure \ref{fig:perms_vs_SAT}, we compare against Qiskit SDK's TokenSwapper algorithm \cite{wagner2023improving} (with 100 trials) for 8-L, 12-O, 27-HH and 65-HH topologies (supplementary section \ref{section:SupplementalInfoCouplingMaps}), and against a SAT solver for 8-L and 12-O. With 100 runs, our algorithm achieves $100\%$ optimality in terms of SWAP count and depth for 8-L and $65\%$ for 12-O. Note that even for 100 runs the execution time for the RL algorithm is under 1 second in total, while on average the SAT solver took 5h for the 12-O permutations. The improvement over TokenSwapper also extends to 27-HH and 65-HH, where the RL model achieves around $50\%$ less SWAP layers.

For Clifford circuits, as shown in Figure \ref{fig:perms_vs_SAT}, we run our model for 100 random Cliffords on 6-Y topology, against a SAT solver and Qiskit SDK's Greedy synthesis algorithm \cite{bravyi2021clifford} (for full connectivity) followed by routing with Qiskit SDK's SABRE. With 1 run, the algorithm takes around the same time as the heuristic algorithm, but achieves a $60\%$ improvement in CNOT layers. With 100 runs (at around 1 second of synthesis time), the results are within 2 CNOT layers from the optimal on average.

\subsection{Circuit Routing with Reinforcement Learning}

The routing task can also be described as a sequential decision making process, similar to the one described in the previous section, but where the possible gates to apply are the SWAPs allowed by the given coupling map, and where instead of an input operator $O_0$ to synthesize, we have an input circuit $C_0^i$ represented in some form, typically a sequence of two-qubit operations. The task then consists on, an each step $t$ and based on the circuit $C_t^i$, selecting the next gate $g_t$ (here only SWAPs) to apply to an output circuit $C_t^o$. When the SWAP is applied, we then check which gates from the circuit $C_t^i$ now fulfill the coupling map (if any), and can ``pass through" to the output circuit $C_t^o$. The process finishes when all the gates have been applied, that is, when we reach an empty circuit $C_T^i$, where $C_T^o$ would contain the routed circuit.

The reward function for training is similar to the one described for synthesis, by providing a large positive reward when a circuit is fully routed, and penalties at each step $t$ based on the two-qubit gate and depth increase produced by action $a_t$ for the output circuit $C_t^o$. For computing this penalty we also allow gate cancellation and mirroring: e.g. applying a SWAP on a pair of qubits for which $C_t^o$ contains a CNOT will incur a lower penalty than applying it on a pair of qubits for which there is no operation to merge with. This encourages the model to learn to leverage existing two-qubit operations as much as possible to route the circuit, generating circuits that might contain more SWAPs a priori but that result in lower two-qubit gate counts after a full transpilation and optimization process.
    
For the layout optimization, we rely on two methods. When using the model for inference, if a SWAP is applied on a pair of qubits that contains no prior operations on $C_t^o$, the SWAP is not applied to $C_t^o$ and instead the initial layout is modified. For further layout optimization we apply bidirectional routing as described in \cite{gushu2019sabre}, where we run multiple iterations of ``forward'' and ``backward'' routing with the RL model and choose the final layout at each step as the initial layout for the next.

We adapt the synthesis algorithm described in the previous section to perform circuit routing with RL. We present two variants of the algorithm: 

\begin{itemize}
    \item A fixed-size routing RL model that performs optimal/near optimal routing for small circuit blocks.
    \item A generic circuit routing RL model, based on SABRE \cite{gushu2019sabre} but with a learned heuristic for SWAP selection, that outperforms the original SABRE heuristic in terms of circuit depth while maintaining and often improving two-qubit gate count.
\end{itemize}

\subsubsection{Fixed size RL routing}
For the fixed-size routing, we encode the full input circuit layer by layer, representing each layer as an $N$ by $N$ binary matrix where non-zero entries mark the two-qubit operations on that layer. We define a maximum horizon $H$ to fix the size of the input to the model. The full circuit is thus encoded as an $N$ by $N$ by $H$ binary tensor. The input to the model is then $N$ by $N$ by $H$ and, since any possible SWAP is allowed at each step, the output is the probability of applying each of the $S$ possible SWAPs of the coupling map. 

\begin{figure*}[tbp]
    \centering
    \begin{minipage}{0.48\textwidth}
        \centering
        \includegraphics[width=1.0\linewidth]{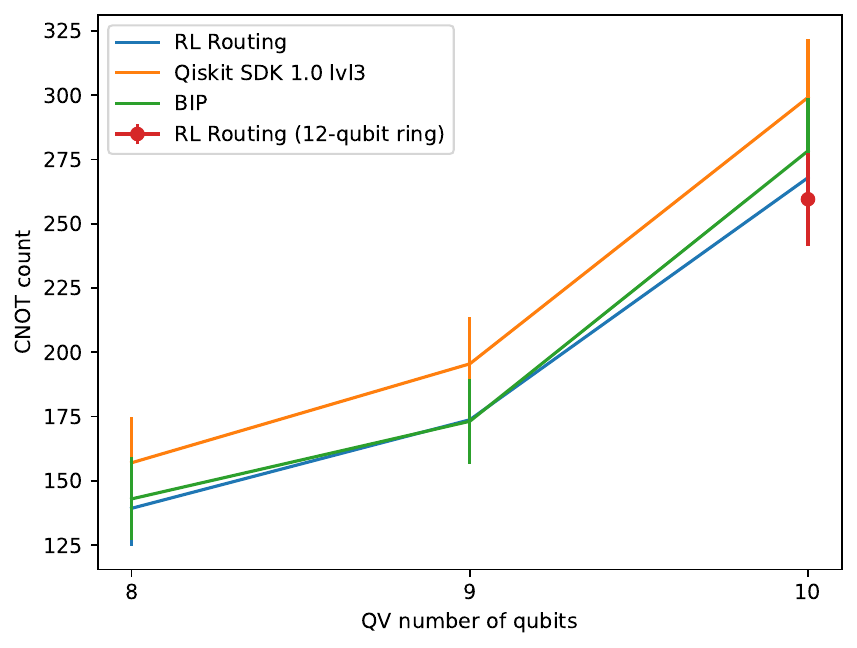} 
    \end{minipage}\hfill
    \begin{minipage}{0.48\textwidth}
        \centering
        \includegraphics[width=1.0\linewidth]{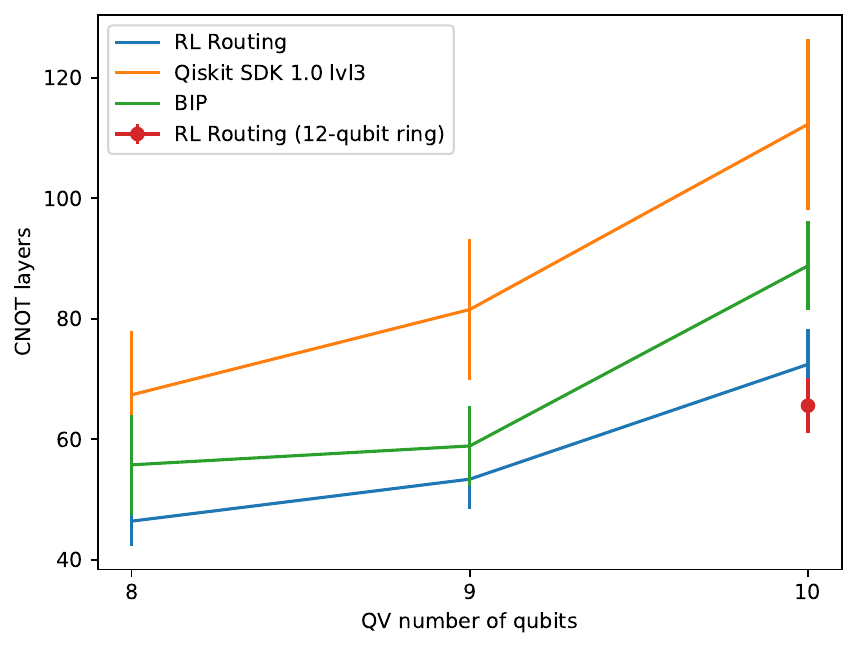}
    \end{minipage}
    \caption{CNOT count (left) and depth (right) for 8-10 qubit quantum volume circuits routed to linear connectivity with different algorithms, and routed to a 12-qubit ring with RL routing.}
    \label{fig:QV_fixed_routing}
\end{figure*}

In Figure \ref{fig:QV_fixed_routing} we show results for routing 8 to 10 qubit quantum volume circuits to linear connectivity, and compare against Qiskit SDK 1.0 level 3 transpiler, and to the BIP mapper method described in \cite{nannicini2022optimal}. As described in the reference, the BIP mapper is used in the ``heuristic'' mode and with a time limit of 10 minutes, so it does not guarantee to produce optimal results. For the RL routing method, we set a time limit of 30 seconds, where we sample routed circuits with the RL method until we reach the time, and post select the best circuit optimizing for CNOT depth first and then CNOT count. The results show that we obtain a $\sim$20\% reduction in CNOT depth with respect to BIP while maintaining and in some cases slightly improving the CNOT count. We also use the RL routing to route the 10 qubit QV circuits into a 12-qubit ring. Interestingly, we see that we can gain a further $\sim$10\% improvement in CNOT depth \textit{and} a $\sim$5\% improvement in CNOT count against routing it to a 10 qubit line.

\subsubsection{Generic RL routing}
For general routing, we follow a strategy similar to SABRE \cite{li2019tackling}, but where we learn the heuristic that selects the next best SWAP with the RL methodology. At each step, we first shortlist the active SWAPs, that is, the SWAPs that are allowed by the coupling map that involve a qubit that is also present in an operation in the front layer of $C_t^i$. We then calculate, for each of the active SWAPs, the variation in distance that each of them would produce on the operations from input circuit, for each of the layers. To fix the size of the input for the model, we fix a horizon $H$ and a max number of active SWAPs $S_a$. The input to the model at each step is then given by an integer matrix of shape $S_a x H$. Because each SWAP involves two qubits, and the circuit $C_t^i$ contains at most two-qubit operations, at each layer, each SWAP can only alter the distances by at most $2$, so the matrix entries are in the range $[2, -2]$.

We run transpiling benchmarks on circuits of up to 133 qubits transpiled for the \device{torino} device \footnote{\device{torino} \url{https://quantum.ibm.com/services/resources?tab=systems&system=ibm_torino} (topology available in supplementary section \ref{section:SupplementalInfoCouplingMaps}).}

\begin{figure*}[tbp]
    \centering
    \begin{minipage}{0.48\textwidth}
        \centering
        \includegraphics[width=1.0\linewidth]{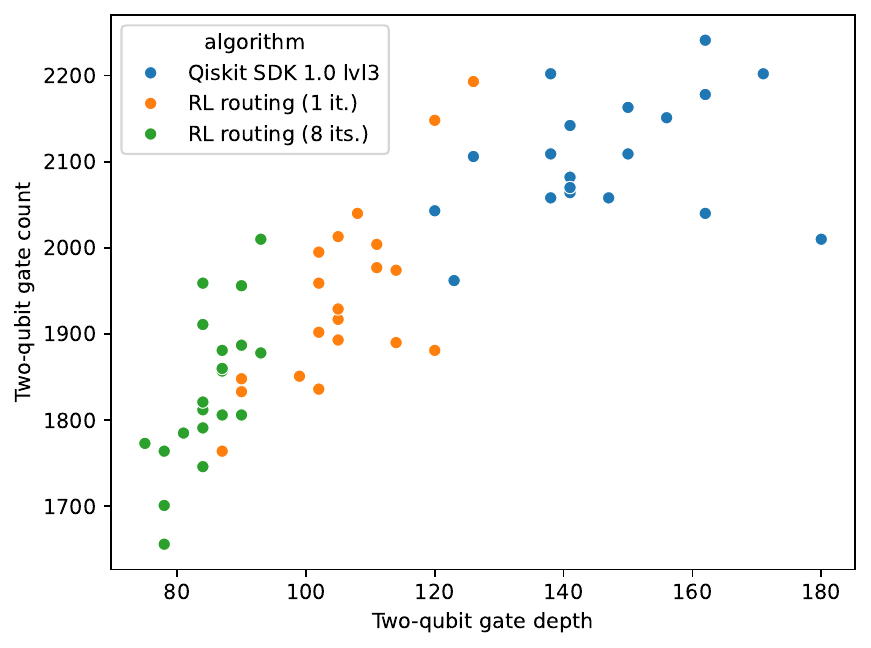}
        \caption{Two-qubit gate count vs two-qubit gate depth for 20 random 3-layer QV circuits on 133 qubits, transpiled to \device{torino}. We run the RL routing algorithm for 1 and 8 iterations.}
        \label{fig:QV_torino_scatter}
    \end{minipage}\hfill
    \begin{minipage}{0.48\textwidth}
        \centering
        \includegraphics[width=1.0\linewidth]{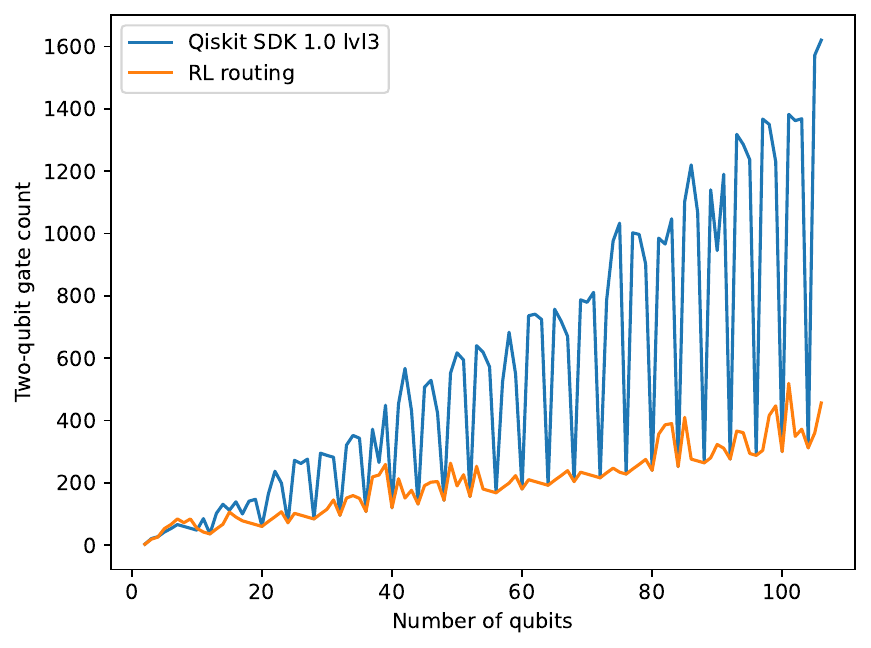}
        \caption{Two-qubit gate count for routing EfficientSU2 circuits with circular entanglement to \device{torino}. We run the RL routing algorithm for 8 iterations using a partial graph matching as the ``seed" layout.}
        \label{fig:SU2_RL_cnot}
    \end{minipage}
\end{figure*}

In Figure~\ref{fig:QV_torino_scatter}, we transpile random 3-layer QV circuits of 133 qubits by running 1 and 8 iterations of the RL routing algorithm. We see that the 8-iteration RL algorithm produces circuits that are on average $\sim$40\% shallower than the Qiskit SDK 1.0 level 3 transpiler, and with $\sim$10\% lower two-qubit gate count. 

In Figure~\ref{fig:SU2_RL_cnot} we transpile EfficientSU2 circuits with circular entanglement for 5 to 110 qubits. We first use a partial matching of the circuit's graph to the connectivity of the device as starting point for the layout, and then run 8 iterations of the RL routing algorithm. The sharp minima on the graph correspond to circuit sizes for which \device{torino}'s coupling map contains a perfect circle and no additional SWAPs need to be inserted. In the regions between the minima, we see that the RL algorithm routes the circuit close to optimally with minimal SWAP overhead.

\tocless\section{Discussion}

As explained in the introduction, the motivation of this work is to enhance the current transpilation workflows with AI with a particular focus on proposing methods that could be practical and efficient. For a variety of transpiling tasks, existing algorithms either suffer from long runtimes or achieve far from optimal results. For practical transpiling workflows, we are interested in getting circuits that are as optimal as possible, but also to do the transpilation within reasonable runtimes. 

There exist examples in literature of various approaches to applying artificial intelligence to quantum circuits. However, the state-of-the-art in this field is still limited. In many cases, existing work relies on existing datasets to train the algorithms  \cite{murakami2022autoqc,arrazola2019machine,acampora2021deep,ACAMPORA2021107526,paler2023machine,quetschlich2023predicting}. Taking as example \cite{bravyi20226}, we observe the creation of a dataset with enough quality and scale (in terms of N of qubits, gates used, etc.) can be extremely large and expensive to compile or generate. This could be a barrier for many problems, as without massive datasets, some AI-based approaches will not be useful or will not scale to solve quantum-circuit-related tasks involving hundreds or even tens of qubits. In contrast, other authors are already exploring the use of other AI-based techniques (RL, genetic algorithms, etc.) where the existence of such datasets is not required \cite{fosel2021quantum, ruiz2024quantum, lockwood2021optimizing, moro2021quantum, arufe2022quantum, rasconi2019innovative, he2021variational,bu2025physical,zen2024quantum,quetschlich2023compiler}. This differentiation is relevant since although standard machine learning algorithms cannot surpass the data-generation process, RL (and similar approaches) demonstrate their strength by learning from experience, and bypass the need for annotated datasets. Continuing with the existing literature, we also observe a lack of device awareness in most of the existing works, meaning they enhance or solve different tasks involving quantum circuits but still require further transpiling to adapt to native instruction sets and connectivity restrictions of most current quantum devices. 

In this paper we have proposed a simple but general framework based on RL that has shown to perform well across different transpiling tasks such as circuit synthesis for different classes of circuits and circuit routing. Our method does not require labeled datasets, and can directly work with native gate sets and connectivity restrictions. The results show that the method is able to push the boundaries of what previous algorithms achieve in terms of the ratio between the quality of results obtained and the runtime of the algorithm, achieving close to optimal results at a fraction of the computational cost of generic optimization methods. 

In the different figures presented through the results (section II) and the tables in the supplementary materials (section  \ref{section:SynthesisBenchmarks}) we demonstrate how our approach is orders of magnitude faster than other traditional approaches that guarantee optimal results (i.e., SAT) and on pair in terms of speed to most of the heuristic algorithms in Qiskit SDK (for $\sim$10 RL runs), while giving better results (optimal or near-optimal) most of the times.  

The combination of quality and speed make this method a strong candidate to be used in a real quantum computing environment. The approach described here is the foundation on which the Qiskit Transpiler Service\footnote{\url{https://docs.quantum.ibm.com/guides/qiskit-transpiler-service}}, and its AI-powered transpiler passes\footnote{\url{https://docs.quantum.ibm.com/guides/ai-transpiler-passes}} are built. The service and its AI features are currently used by IBM Quantum users in real quantum computing workflows with positive results and a significant reduction of CNOTs and layers of CNOTs for most of the circuits. 

We have shown different results for each type of circuit. For Linear Function circuits we present results up to 9 qubits, for Clifford up to 11 qubits, for Permutation up to 65 qubits and we discuss how we run routing up to 133 qubits. Each type of circuit has their own actions space and particularities, and our approach scales differently for them. We acknowledge that the synthesis methods, as presented here, could struggle to scale to thousands of qubits, as we have seen that the training procedure takes longer to converge when increasing the circuit size. As far as we know, there is no fundamental limitation on scaling the method -the size of the networks might in principle scale linearly with the size of the operators and the gate sets-, so improving the method to train more efficiently on larger circuits is an interesting research direction. However, we argue that, in practice, the method is useful even with the circuit sizes presented here, since we use this as an optimization stage for re-synthesizing certain blocks within larger circuits. These statements presented about the potential of our solution with RL in terms of scalability, applicability in real contexts, and competitiveness in the balance of efficiency and results have been independently verified in studies such as \cite{nation2024benchmarking} where the authors have been able to execute benchmarks that show how our AI methods included in the Qiskit Transpiler Service are able to scale to circuits of up to 900 qubits, and one million two-qubit gates, yielding results that are at least $\sim$30\% better in terms of depth and two-qubit gates count on average compared to other quantum computing libraries and their heuristic algorithms.

Comparing to other RL-based methods already published, we see our approach can be improved by adding procedures like Monte Carlo Tree Search for exploring the action space more effectively. One disadvantage we see of that kind of strategies is that it may result on longer runtimes, and may be less useful if one wants a reasonably good solution as fast as possible. Another interesting direction would be to train generic models: as discussed earlier, we are currently training different RL models for different connectivity graphs or topologies. This means that in practice we need to have models for every possible topology we want to do synthesis for. Instead, this could also be achieved by having a generic model that contains all the possible gates in its action set, and use action masking for training and inference depending on the specific restrictions we want to impose. 

We see that the method can be easily applied to other families of circuits and to other problems found on typical transpiling pipelines. One particular area of interest is the synthesis of dynamic circuits, that is, circuits that contain non-unitary operations such as measurements, resets and conditional gates. Since this is a relatively unexplored area, the RL algorithm might be helpful in discovering interesting circuit identities that leverage dynamic operations to optimize circuits further. We also see that this can be applied to generate specialized routing algorithms for specific types of circuits of interest, resulting in better routing that what could be achieved with generic routing algorithms. Finally, we believe this method can also be applied to constrained Pauli network synthesis (synthesis of sequences of Pauli rotations), where there are algorithms for full connectivity but no efficient methods for synthesis on native device instruction sets with connectivity constraints.

\vspace{2em}
\tocless{\section*{Author contributions}}
Conceptualization: D. K., I. D., I. F., J. C-B. Technical implementation: D. K., V. V., J. C-B. Training of RL algorithms: D. K., V. V., J. C-B. Execution of benchmarks: D. K., V. V. All authors contributed to the manuscript writing.

\vspace{2em}
\tocless\acknowledgements
We would like to express our gratitude to the Quantum+AI team at IBM Quantum for their efforts in implementing the procedures presented in this paper within the Qiskit Transpiler Service. We specifically acknowledge Jesús Talavera, Sanjay Vishwakarma, and Yaiza García Martín Mantero for their invaluable contributions to this project. The authors acknowledge feedback and insightful discussions with Ali Javadi, Blake R. Johnson, Lev Bishop, Jay Gambetta, and Francisco J. Martín-Fernández. The authors also acknowledge the IBM Research Cognitive Computing Cluster service for providing resources that have contributed to the research results reported within this paper.

\tocless\bibliography{references}

\clearpage
\setcounter{section}{0}  
\setcounter{figure}{0}   
\renewcommand{\thefigure}{S\arabic{figure}}

\onecolumngrid
\setcounter{page}{1}

\tocless{\section*{Supplementary Information: Practical and efficient quantum circuit synthesis and transpiling with Reinforcement Learning}}

\subsection{Technical details for training}

To train our RL agents, we utilize the Proximal Policy Optimization (PPO) algorithm \cite{schulman2017proximal} as implemented in Stable Baselines3 \cite{stable-baselines3}, serving as our baseline method. In conjunction with custom RL environments, this approach enables our agents to learn how to tackle various tasks while interacting with different types of circuits and their associated options.

The PPO algorithm is an RL on-policy method that alternates between sampling data through interacting with the environment and optimizing a surrogate objective function using stochastic gradient ascent \cite{schulman2017proximal}. It means that the PPO algorithm updates its policy as it interacts with the environment so that policy changes are incremental (avoiding extensive policy updates via clipping), improving the stability of agent training. 

The custom RL environments are the core software module that shape how agents can resolve the desired tasks (routing, synthesis) while following some software patterns and interfaces that make it usable while training using the PPO algorithm. The environments are implemented following the OpenAI Gym \cite{openaigym} interface and include utilities also from RLlib \cite{liang2018rllib}. The environments are responsible for providing general methods to the agents to choose the following action to perform, get observations from actions, reset the environment after each episode, define seeds for random actions, or manage running multiple vectorized environments in parallel to speed up the training. Prominent features of these custom RL environments include the ability to generate customizable circuits (depth, number of qubits, ratio of CNOT gates vs. other gates, randomness of the circuits), adjust circuit difficulty, represent circuits in various ways, and restrict qubit connections to match specific quantum device layouts. The environments leverage Qiskit SDK methods \cite{Qiskit,javadi2024quantum} to generate, transform, and manage various types of circuits. Lastly, the environments are capable of computing the reward function using various factors as inputs, such as the number of CNOTs in synthesized circuits, the total number of gates, the number of layers, or the number of layers containing CNOT gates.

Throughout the training process, we acknowledge the use of the Optuna framework \cite{optuna2019} to optimize the hyperparameters employed to configure the PPO algorithm.

The RL models used for this paper were trained in machines with the following hardware specifications. Platform system: Linux-4.18.0-372.26.1.el8\_6.x86\_64-x86\_64-with-glibc2.28, Python version: 3.9.1, available RAM : 1007.316 GB, CPU count: 128, CPU model: AMD EPYC 7742 64-Core Processor, GPU: 1 x NVIDIA A100-SXM4-40GB. The training time for all the different models reported in this paper varies from minutes to less than 48 hours using that hardware setup.

\subsection{Synthesis benchmarks}\label{section:SynthesisBenchmarks}
Tables \ref{tab:results_summary_permutation}, \ref{tab:results_summary_linear_function} and \ref{tab:results_summary_linear_function} present a compelling summary of the results obtained for the different synthesis benchmarks over the different types of circuits and topologies tested.

\begin{table}[h]
\caption{Benchmarks for Permutation synthesis comparing our RL (1, 10, 100 and 1000 runs) vs TokenSwapper permutation synthesis method \cite{wagner2023improving} with 100 trials. The results are for 100 random permutations for each number of qubits. The same permutations are used for each number of qubits, so the results also show how well the different topologies can accommodate permutations on average. For reference about the topologies, review supplementary section \ref{section:SupplementalInfoCouplingMaps}}
\label{tab:results_summary_permutation}
\begin{tabular}{l||rrrrr|rrrrr|rrrrr}
\toprule
  & \multicolumn{5}{c}{time (miliseconds)} & \multicolumn{5}{c}{SWAP count} & \multicolumn{5}{c}{SWAP layers} \\
algorithm & TS$_{100}$ & RL$_{1}$ & RL$_{10}$ & RL$_{100}$ & RL$_{1000}$ & TS$_{100}$ & RL$_{1}$ & RL$_{10}$ & RL$_{100}$ & RL$_{1000}$ & TS$_{100}$ & RL$_{1}$ & RL$_{10}$ & RL$_{100}$ & RL$_{1000}$ \\
\hline
topology &  &  &  &  &  &  &  &  &  &  &  &  &  &  &  \\
\midrule
8-L & 16.1 & 6.0 & 23.3 & 123.0 & 1142.0 & 13.5 & 13.5 & 13.5 & 13.5 & 13.5 & 8.4 & 6.1 & 5.8 & 5.8 & 5.8 \\
12-O & 41.3 & 15.1 & 53.1 & 445.4 & 4306.4 & 25.0 & 24.0 & 23.1 & 22.7 & 22.4 & 11.3 & 8.2 & 7.5 & 7.1 & 6.8 \\
27-HH & 222.8 & 12.7 & 80.2 & 776.4 & 7632.4 & 87.3 & 81.0 & 79.0 & 78.3 & 78.0 & 32.6 & 20.0 & 18.4 & 17.3 & 16.8 \\
33-HH & 345.9 & 17.8 & 119.9 & 1135.1 & 11253.8 & 122.3 & 118.5 & 115.8 & 114.2 & 113.0 & 41.2 & 26.0 & 24.2 & 22.9 & 22.5 \\
65-HH & 154.1 & 66.7 & 410.3 & 3884.4 & 36654.4 & 362.4 & 353.9 & 344.8 & 340.7 & 336.3 & 92.1 & 51.4 & 48.9 & 47.5 & 46.2 \\
\bottomrule
\end{tabular}
\end{table}

\begin{table}[h]
\caption{Benchmarks for Linear Function synthesis comparing our RL (1, 10, 100 and 1000 runs) vs Patel-Markov-Hayes (PMH) Linear Function synthesis method \cite{markov2008optimal} followed by SABRE routing. Note that the PMH + SABRE circuits implement the Linear Functions up to a final permutation due to the routing, while the RL one preserves the permutation; the PMH CNOT count and depth will be higher if we included these. The results are for 100 random Linear Functions for each number of qubits. The same Linear Functions are used for each number of qubits, so the results also show how well the different topologies can accommodate Linear Functions on average. For reference about the topologies, review supplementary section \ref{section:SupplementalInfoCouplingMaps}}
\label{tab:results_summary_linear_function}
\begin{tabular}{l||rrrrr|rrrrr|rrrrr}
\toprule
 & \multicolumn{5}{c}{time (miliseconds)} & \multicolumn{5}{c}{CNOT count} & \multicolumn{5}{c}{CNOT layers} \\
algorithm & PMH & RL$_{1}$ & RL$_{10}$ & RL$_{100}$ & RL$_{1000}$ & PMH & RL$_{1}$ & RL$_{10}$ & RL$_{100}$ & RL$_{1000}$ & PMH & RL$_{1}$ & RL$_{10}$ & RL$_{100}$ & RL$_{1000}$ \\
\hline
topology &  &  &  &  &  &  &  &  &  &  &  &  &  &  &  \\
\midrule
3-L & 16.5 & 1.5 & 5.6 & 47.4 & 451.1 & 7.2 & 4.8 & 4.8 & 4.8 & 4.8 & 7.2 & 4.8 & 4.8 & 4.8 & 4.8 \\
\hline
4-L & 20.0 & 2.5 & 11.4 & 100.1 & 983.5 & 15.6 & 10.2 & 10.1 & 10.0 & 9.9 & 13.9 & 8.9 & 8.7 & 8.7 & 8.6 \\
4-Y & 21.8 & 2.4 & 9.9 & 86.4 & 853.0 & 12.9 & 8.3 & 8.2 & 8.1 & 8.1 & 12.9 & 8.3 & 8.2 & 8.1 & 8.1 \\
\hline
5-L & 18.6 & 4.0 & 19.9 & 179.5 & 1768.8 & 29.9 & 17.2 & 16.4 & 16.1 & 15.9 & 24.5 & 13.8 & 12.8 & 12.8 & 12.6 \\
5-T & 20.8 & 3.6 & 17.5 & 157.7 & 1570.1 & 24.8 & 14.8 & 14.2 & 13.9 & 13.7 & 21.9 & 12.7 & 12.3 & 12.1 & 11.9 \\
\hline
6-L & 22.2 & 5.3 & 31.2 & 291.4 & 2923.6 & 53.3 & 27.1 & 25.9 & 25.4 & 25.0 & 40.1 & 19.3 & 18.5 & 18.1 & 17.8 \\
6-T & 23.0 & 5.1 & 29.2 & 265.3 & 2605.9 & 45.8 & 23.9 & 23.1 & 22.5 & 22.1 & 36.9 & 18.5 & 17.9 & 17.3 & 16.9 \\
6-Y & 25.0 & 4.6 & 27.6 & 253.1 & 2541.1 & 44.4 & 23.1 & 22.2 & 21.6 & 21.3 & 35.8 & 18.5 & 17.8 & 17.2 & 17.0 \\
\hline
7-F & 25.6 & 6.9 & 44.1 & 409.6 & 4010.0 & 72.1 & 35.8 & 34.0 & 33.1 & 32.5 & 53.7 & 25.0 & 23.9 & 22.8 & 22.7 \\
7-H & 23.5 & 6.7 & 39.9 & 369.6 & 3645.2 & 66.4 & 33.1 & 31.4 & 30.6 & 30.1 & 50.7 & 24.1 & 23.1 & 22.3 & 21.9 \\
7-L & 24.7 & 7.7 & 47.3 & 450.9 & 4474.5 & 84.3 & 40.1 & 38.5 & 37.5 & 36.9 & 58.5 & 25.3 & 24.4 & 23.9 & 23.6 \\
7-T & 22.1 & 6.7 & 43.4 & 411.3 & 4088.3 & 76.2 & 36.7 & 35.1 & 34.3 & 33.9 & 55.4 & 24.6 & 23.2 & 22.9 & 22.2 \\
7-Y & 23.2 & 6.5 & 41.1 & 389.7 & 3789.6 & 67.9 & 34.4 & 32.0 & 31.0 & 30.4 & 51.9 & 24.4 & 22.8 & 22.4 & 22.4 \\
\hline
8-F & 27.0 & 9.3 & 61.1 & 584.3 & 5772.9 & 116.3 & 52.2 & 49.2 & 47.6 & 46.7 & 79.4 & 31.8 & 30.6 & 29.2 & 28.7 \\
8-H & 25.8 & 9.1 & 58.2 & 546.2 & 5338.3 & 104.2 & 48.9 & 46.4 & 45.0 & 43.8 & 74.6 & 31.6 & 29.9 & 29.0 & 28.6 \\
8-L & 26.9 & 11.3 & 72.2 & 671.3 & 6592.0 & 135.8 & 58.4 & 55.2 & 53.9 & 52.8 & 86.6 & 32.1 & 30.7 & 30.4 & 29.5 \\
8-T1 & 25.5 & 10.0 & 62.8 & 586.7 & 5813.5 & 123.5 & 54.1 & 51.0 & 49.5 & 48.6 & 80.8 & 31.9 & 29.9 & 29.1 & 28.7 \\
8-T2 & 25.7 & 9.0 & 59.9 & 573.8 & 5662.5 & 106.3 & 50.6 & 47.2 & 45.4 & 44.3 & 74.7 & 31.9 & 29.9 & 28.4 & 28.6 \\
8-Y & 27.3 & 9.6 & 62.0 & 575.7 & 5784.7 & 113.2 & 52.7 & 49.7 & 47.7 & 46.3 & 78.1 & 32.0 & 30.4 & 29.2 & 29.2 \\
\hline
9-F1 & 31.5 & 13.1 & 87.5 & 848.8 & 8487.4 & 180.7 & 70.5 & 67.0 & 64.8 & 63.2 & 111.3 & 38.6 & 36.5 & 35.4 & 34.4 \\
9-F2 & 27.8 & 13.1 & 88.8 & 862.9 & 8533.0 & 175.8 & 71.7 & 66.9 & 64.6 & 63.2 & 111.5 & 40.1 & 37.2 & 36.2 & 35.1 \\
9-H1 & 29.1 & 12.4 & 81.3 & 780.9 & 7785.4 & 176.1 & 67.8 & 63.7 & 61.9 & 60.8 & 111.2 & 38.2 & 35.2 & 34.2 & 32.9 \\
9-H2 & 31.8 & 12.2 & 78.2 & 762.1 & 7490.1 & 157.0 & 65.1 & 60.5 & 58.3 & 57.0 & 105.1 & 38.0 & 35.5 & 34.1 & 33.0 \\
9-H3 & 29.1 & 11.3 & 78.6 & 756.9 & 7472.3 & 152.4 & 65.4 & 60.8 & 58.6 & 57.1 & 103.3 & 39.7 & 36.5 & 35.4 & 34.4 \\
9-H4 & 32.0 & 11.7 & 79.5 & 753.9 & 7537.9 & 159.5 & 65.6 & 61.4 & 59.3 & 57.7 & 104.7 & 38.9 & 36.0 & 34.9 & 34.3 \\
9-L & 32.0 & 14.3 & 99.1 & 912.3 & 9092.8 & 208.2 & 79.1 & 73.8 & 71.1 & 69.5 & 125.5 & 40.7 & 37.4 & 36.4 & 35.8 \\
9-T1 & 30.9 & 13.6 & 90.4 & 880.5 & 8748.5 & 194.3 & 74.6 & 69.8 & 67.5 & 66.0 & 119.7 & 39.5 & 36.6 & 35.2 & 34.1 \\
9-T2 & 30.0 & 12.1 & 83.1 & 780.0 & 7927.2 & 169.8 & 69.8 & 64.4 & 61.4 & 59.9 & 110.1 & 39.6 & 36.3 & 34.9 & 34.1 \\
9-Y & 27.7 & 12.3 & 85.2 & 801.0 & 7951.0 & 161.5 & 69.9 & 64.3 & 61.6 & 59.5 & 107.4 & 39.7 & 36.9 & 35.5 & 34.1 \\
\bottomrule
\end{tabular}
\end{table}

\begin{table}[h]
\caption{Benchmarks for Clifford synthesis comparing our RL (1, 10, 100 and 1000 runs) vs Qiskit SDK's Greedy Clifford synthesis method \cite{bravyi2021clifford} followed by SABRE routing. Note that the Greedy + SABRE circuits implement the Cliffords up to a final permutation due to the routing, while the RL one preserves the permutation; the Greedy CNOT count and depth will be higher if we included these. The results are for 100 random Cliffords for each number of qubits. The same Cliffords are used for each number of qubits, so the results also show how well the different topologies can accommodate Cliffords on average. For reference about the topologies, review supplementary section \ref{section:SupplementalInfoCouplingMaps}}
\label{tab:results_summary_clifford}
\begin{tabular}{l||rrrrr|rrrrr|rrrrr}
\toprule
 & \multicolumn{5}{c}{time (miliseconds)} & \multicolumn{5}{c}{CNOT count} & \multicolumn{5}{c}{CNOT layers} \\
algorithm & Greedy & RL$_{1}$ & RL$_{10}$ & RL$_{100}$ & RL$_{1000}$ & Greedy & RL$_{1}$ & RL$_{10}$ & RL$_{100}$ & RL$_{1000}$ & Greedy & RL$_{1}$ & RL$_{10}$ & RL$_{100}$ & RL$_{1000}$ \\
\hline
topology &  &  &  &  &  &  &  &  &  &  &  &  &  &  &  \\
\midrule
3-L & 28.9 & 13.1 & 23.6 & 145.2 & 1356.9 & 6.5 & 5.0 & 4.7 & 4.7 & 4.6 & 6.5 & 5.0 & 4.7 & 4.7 & 4.6 \\
\hline
4-L & 28.1 & 17.2 & 41.5 & 300.5 & 2891.8 & 23.7 & 10.2 & 9.4 & 9.1 & 8.9 & 21.5 & 9.0 & 7.8 & 7.4 & 7.2 \\
4-T & 27.4 & 16.2 & 41.1 & 287.5 & 2714.9 & 19.0 & 9.4 & 8.5 & 8.2 & 8.0 & 19.0 & 9.4 & 8.5 & 8.2 & 8.0 \\
\hline
5-L & 35.8 & 50.3 & 293.2 & 2650.5 & 26987.5 & 44.2 & 16.6 & 15.4 & 14.7 & 14.4 & 36.4 & 13.1 & 11.7 & 10.6 & 10.0 \\
5-T & 34.0 & 47.1 & 279.4 & 2669.4 & 26854.3 & 36.3 & 16.8 & 14.6 & 13.6 & 13.2 & 32.5 & 14.8 & 12.5 & 11.6 & 11.1 \\
\hline
6-L & 41.2 & 46.0 & 153.4 & 1220.2 & 12068.1 & 72.0 & 25.9 & 23.9 & 23.1 & 22.6 & 55.9 & 18.2 & 16.0 & 14.8 & 14.2 \\
6-T & 40.0 & 43.6 & 147.1 & 1187.9 & 11295.5 & 61.8 & 23.9 & 22.4 & 21.5 & 21.0 & 50.4 & 18.3 & 16.4 & 15.4 & 14.8 \\
6-Y & 40.8 & 40.6 & 139.4 & 1126.5 & 11094.3 & 59.2 & 22.7 & 20.9 & 19.9 & 19.4 & 49.6 & 17.9 & 16.1 & 15.0 & 14.4 \\
\hline
7-H & 51.0 & 150.2 & 1095.3 & 10579.2 & 105140.2 & 88.8 & 35.5 & 31.3 & 29.3 & 27.9 & 69.3 & 25.7 & 21.8 & 20.3 & 18.8 \\
7-L & 50.6 & 72.8 & 277.7 & 2320.0 & 22892.2 & 111.4 & 38.5 & 35.5 & 34.4 & 33.6 & 80.1 & 24.6 & 22.2 & 20.4 & 19.4 \\
7-T & 50.4 & 68.1 & 260.1 & 2183.2 & 21725.6 & 98.0 & 36.4 & 33.7 & 32.3 & 31.5 & 73.5 & 26.7 & 23.3 & 21.8 & 20.8 \\
7-F & 50.7 & 66.6 & 262.5 & 2193.7 & 21637.9 & 92.0 & 34.6 & 32.4 & 31.2 & 30.2 & 71.4 & 24.4 & 21.9 & 20.5 & 19.7 \\
7-Y & 56.4 & 65.3 & 254.0 & 2112.8 & 21117.8 & 88.4 & 33.5 & 30.6 & 29.0 & 28.0 & 70.0 & 25.2 & 22.5 & 20.6 & 19.8 \\
\hline
8-H & 60.0 & 104.4 & 537.5 & 4748.2 & 45816.2 & 126.9 & 44.6 & 44.4 & 41.9 & 40.4 & 94.5 & 29.8 & 29.1 & 26.7 & 25.3 \\
8-L & 64.8 & 110.4 & 528.2 & 4679.2 & 46280.0 & 159.7 & 51.3 & 49.3 & 47.3 & 46.0 & 108.8 & 30.9 & 28.7 & 27.0 & 25.6 \\
8-T & 60.7 & 109.6 & 515.4 & 4500.8 & 44844.4 & 145.3 & 50.2 & 47.2 & 44.8 & 43.5 & 101.4 & 32.8 & 29.8 & 27.6 & 26.4 \\
8-T2 & 60.7 & 102.5 & 498.7 & 4468.2 & 45043.6 & 129.5 & 47.5 & 45.8 & 43.4 & 42.1 & 95.5 & 34.2 & 32.6 & 30.7 & 28.8 \\
8-F & 62.4 & 106.8 & 501.4 & 4482.2 & 44533.8 & 137.3 & 48.2 & 46.5 & 44.4 & 43.0 & 97.8 & 31.5 & 29.6 & 27.6 & 26.0 \\
8-Y & 62.7 & 113.3 & 468.8 & 4390.9 & 44764.2 & 138.5 & 49.0 & 44.8 & 42.9 & 41.5 & 100.7 & 32.0 & 28.1 & 26.4 & 25.0 \\
\hline
9-H1 & 78.2 & 183.8 & 943.7 & 8243.7 & 82258.1 & 198.2 & 86.2 & 76.9 & 71.3 & 68.1 & 133.7 & 49.3 & 43.7 & 40.3 & 37.7 \\
9-H2 & 76.2 & 155.3 & 780.0 & 6613.8 & 67336.5 & 177.6 & 70.8 & 65.5 & 61.9 & 59.7 & 125.9 & 42.9 & 39.5 & 37.0 & 35.1 \\
9-H3 & 79.7 & 168.4 & 897.3 & 7798.2 & 74386.3 & 172.1 & 71.8 & 64.8 & 61.4 & 58.9 & 123.0 & 44.1 & 39.8 & 37.3 & 35.8 \\
9-H4 & 76.1 & 167.3 & 854.5 & 7439.1 & 69722.8 & 180.9 & 68.6 & 63.6 & 60.6 & 58.5 & 125.4 & 46.5 & 42.5 & 40.0 & 38.1 \\
9-L & 75.2 & 175.2 & 865.4 & 7665.9 & 76943.3 & 227.1 & 83.8 & 77.0 & 73.0 & 70.7 & 145.1 & 45.8 & 41.5 & 38.9 & 37.8 \\
9-T & 74.4 & 179.4 & 896.2 & 7711.2 & 74739.2 & 208.9 & 79.0 & 72.0 & 68.7 & 66.0 & 137.9 & 47.1 & 41.9 & 39.8 & 37.2 \\
9-T2 & 76.6 & 165.4 & 848.3 & 7593.8 & 73196.4 & 190.0 & 69.5 & 64.4 & 61.7 & 59.5 & 131.8 & 44.6 & 41.0 & 38.4 & 36.8 \\
9-F & 76.5 & 182.4 & 972.0 & 9021.2 & 87831.0 & 206.7 & 71.3 & 73.9 & 70.6 & 68.8 & 137.8 & 40.1 & 41.1 & 38.4 & 37.0 \\
9-F2 & 77.2 & 190.7 & 1246.3 & 10838.0 & 107162.8 & 197.9 & 65.5 & 79.1 & 76.0 & 73.5 & 133.4 & 35.8 & 41.8 & 39.2 & 37.4 \\
9-Y & 74.7 & 186.3 & 1062.2 & 9578.1 & 94713.1 & 183.9 & 60.7 & 60.2 & 57.8 & 55.5 & 129.3 & 38.0 & 37.4 & 35.3 & 33.4 \\

\hline
10-L & 97.9 & 226.8 & 1289.0 & 11373.0 & 109692.0 & 302.4 & 100.8 & 94.0 & 90.0 & 87.5 & 186.0 & 50.2 & 46.6 & 43.4 & 42.3 \\
\hline
11-L & 117.1 & 393.7 & 2375.9 & 21431.4 & 199072.4 & 402.8 & 140.8 & 130.6 & 123.3 & 118.8 & 233.3 & 71.3 & 63.8 & 59.8 & 57.4 \\

\bottomrule
\end{tabular}
\end{table}

\subsection{Reference of coupling maps used in the paper}\label{section:SupplementalInfoCouplingMaps}

This section includes every coupling map used in this paper with their different related topology IDs (as reported in previous sections).

\begin{table}[htbp]
    \centering
    \label{tab:images}
        \begin{tabular}{c c c}
            {\includegraphics[width=0.15\columnwidth,valign=m]{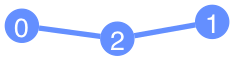} \label{fig:topology_3qL}}& 
            {\includegraphics[width=0.25\columnwidth,valign=m]{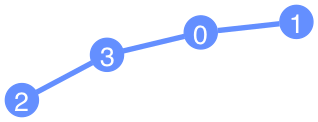}\label{fig:topology_4qL}} &
            {\includegraphics[width=0.18\columnwidth,valign=m]{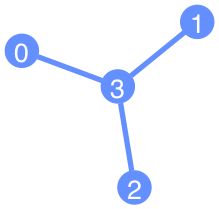} \label{fig:topology_4qY}} \\
            Topology ID: 3-L & Topology ID: 4-L & Topology ID: 4-Y\\
            &&\\
            {\includegraphics[height=0.25\columnwidth,valign=m]{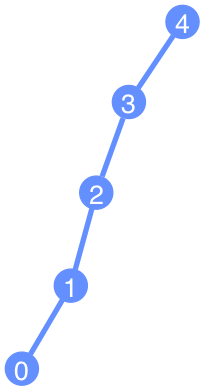} \label{fig:topology_5qL}} &{\includegraphics[height=0.25\columnwidth,valign=m]{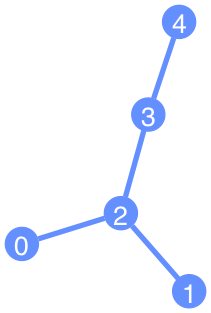} \label{fig:topology_5qT}} &
            {\includegraphics[height=0.25\columnwidth,valign=m]{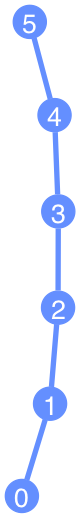} \label{fig:topology_6qL}} \\
            Topology ID: 5-L & Topology ID: 5-T& Topology ID: 6-L \\
            &&\\
            {\includegraphics[height=0.25\columnwidth,valign=m]{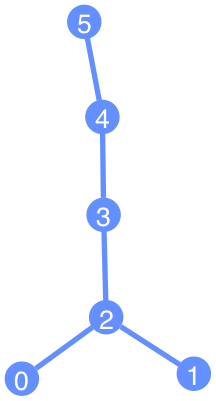} \label{fig:topology_6qT}} &{\includegraphics[height=0.25\columnwidth,valign=m]{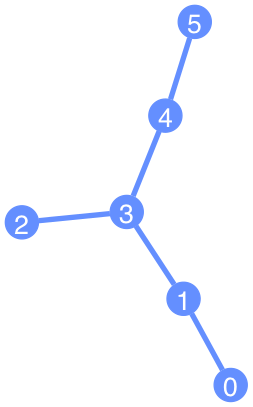} \label{fig:topology_6qY}} &
            {\includegraphics[width=0.25\columnwidth,valign=m]{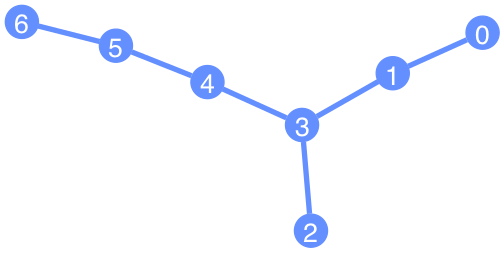} \label{fig:topology_7qF}} \\
            Topology ID: 6-T & Topology ID: 6-Y& Topology ID: 7-F \\
            &&\\
            {\includegraphics[height=0.25\columnwidth,valign=m]{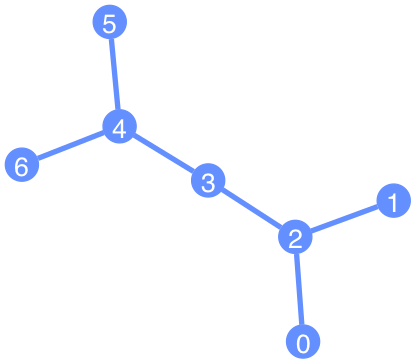} \label{fig:topology_7qH}} &{\includegraphics[height=0.25\columnwidth,valign=m]{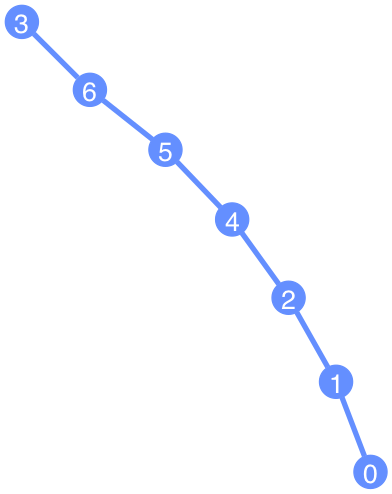} \label{fig:topology_7qL}} &
            {\includegraphics[width=0.25\columnwidth,valign=m]{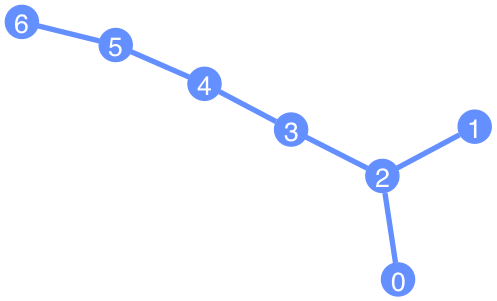} \label{fig:topology_7qT}} \\
            Topology ID: 7-H & Topology ID: 7-L& Topology ID: 7-T \\

        \end{tabular}
\end{table}

\begin{table}[htbp]
    \centering
    \label{tab:images}
        \begin{tabular}{c c c}
            {\includegraphics[width=0.15\columnwidth,valign=m]{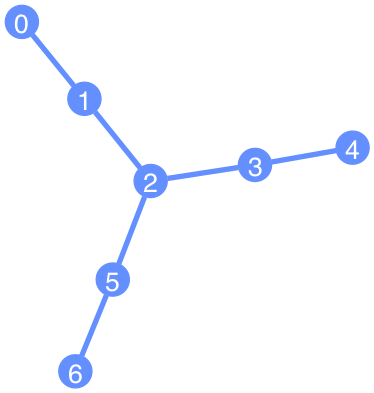} \label{fig:topology_7qY}}& 
            {\includegraphics[height=0.25\columnwidth,valign=m]{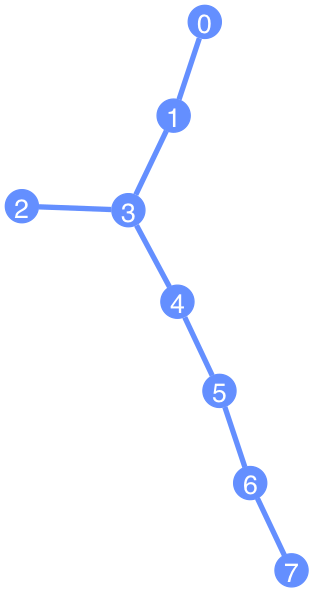}\label{fig:topology_8qF}} &
            {\includegraphics[width=0.18\columnwidth,valign=m]{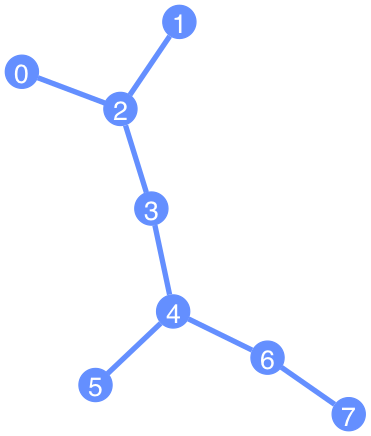} \label{fig:topology_8qJ}} \\
            Topology ID: 7-Y & Topology ID: 8-F & Topology ID: 8-H\\
            &&\\
            {\includegraphics[height=0.25\columnwidth,valign=m]{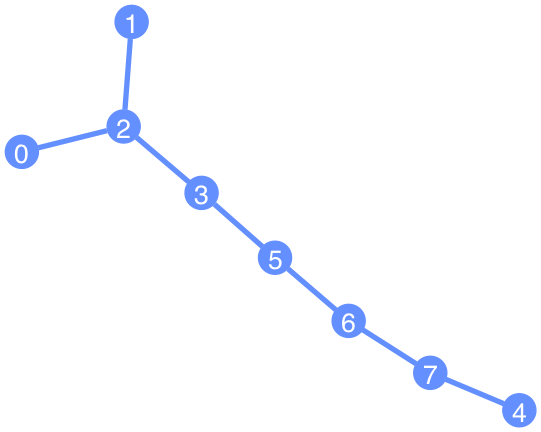} \label{fig:topology_8qT1}} &{\includegraphics[height=0.25\columnwidth,valign=m]{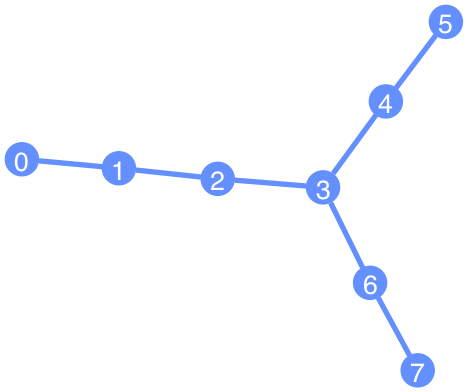} \label{fig:topology_8qT2}} &
            {\includegraphics[height=0.25\columnwidth,valign=m]{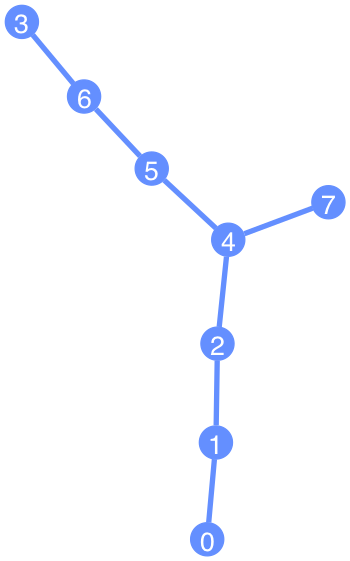} \label{fig:topology_8qY}} \\
            Topology ID: 8-T1 & Topology ID: 8-T2& Topology ID: 8-Y \\
            &&\\
            {\includegraphics[height=0.25\columnwidth,valign=m]{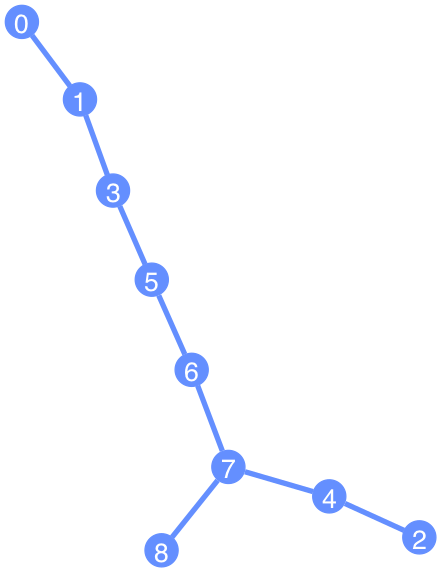} \label{fig:topology_9qF1}} &{\includegraphics[height=0.25\columnwidth,valign=m]{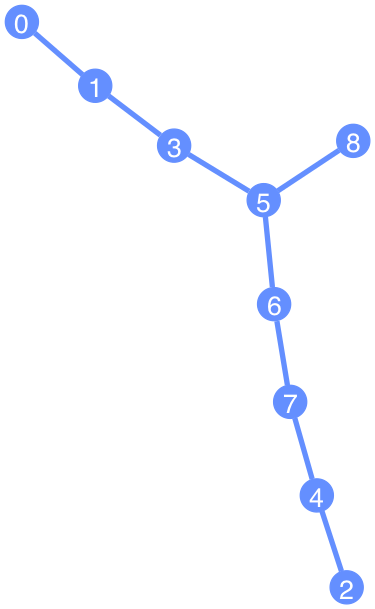} \label{fig:topology_9qF2}} &
            {\includegraphics[width=0.25\columnwidth,valign=m]{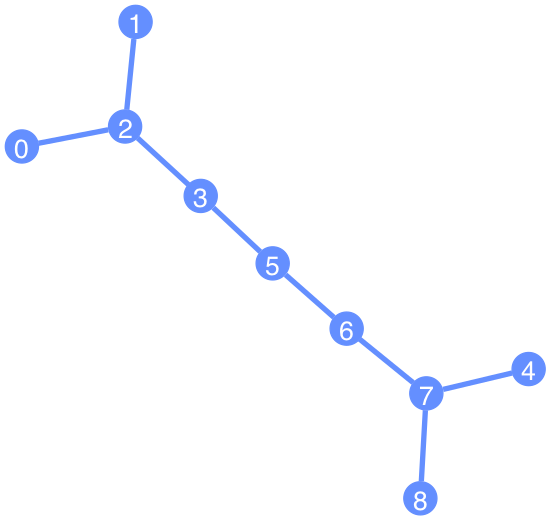} \label{fig:topology_9qH1}} \\
            Topology ID: 9-F1 & Topology ID: 9-F2& Topology ID: 9-H1 \\
            &&\\
            {\includegraphics[height=0.25\columnwidth,valign=m]{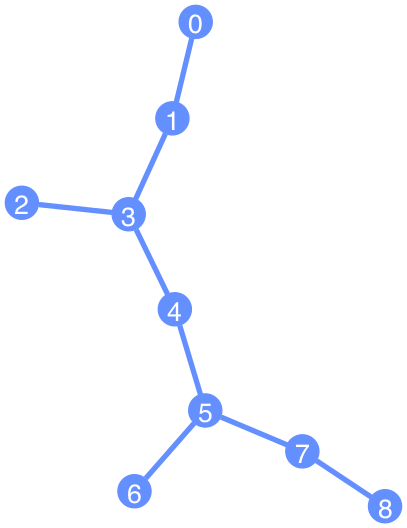} \label{fig:topology_9qH2}} &{\includegraphics[height=0.25\columnwidth,valign=m]{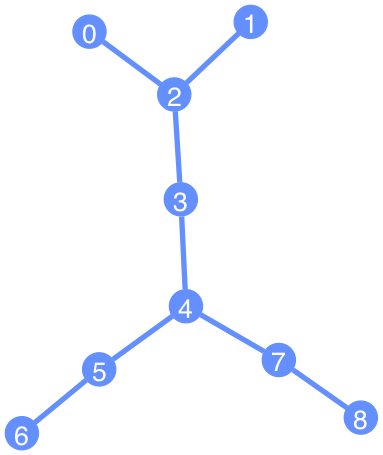} \label{fig:topology_9qH3}} &
            {\includegraphics[width=0.25\columnwidth,valign=m]{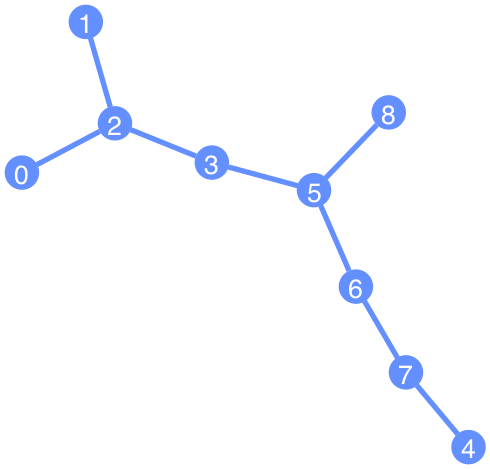} \label{fig:topology_9qJH4}} \\
            Topology ID: 9-H2 & Topology ID: 9-H3& Topology ID: 9-H4 \\

        \end{tabular}
\end{table}

\begin{table}[htbp]
    \centering
    \label{tab:images}
        \begin{tabular}{c c c}
            {\includegraphics[height=0.2\columnwidth,valign=m]{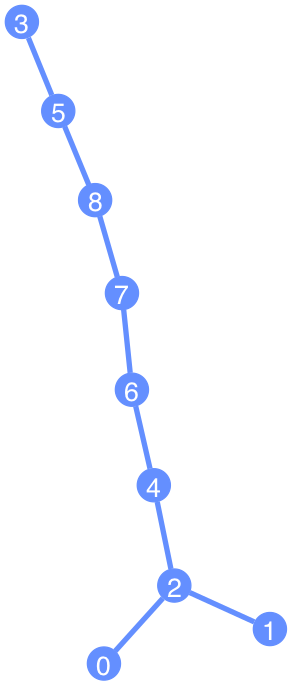} \label{fig:topology_9qT1}}& 
            {\includegraphics[height=0.25\columnwidth,valign=m]{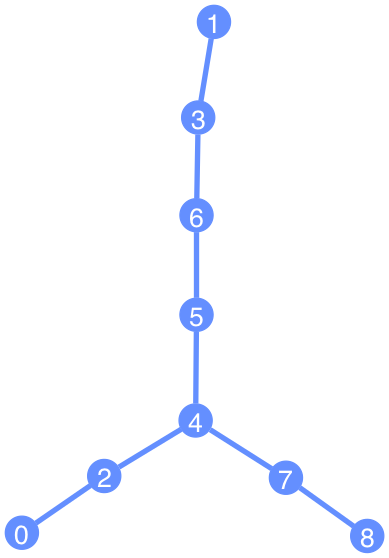}\label{fig:topology_9qT2}} &
            {\includegraphics[width=0.18\columnwidth,valign=m]{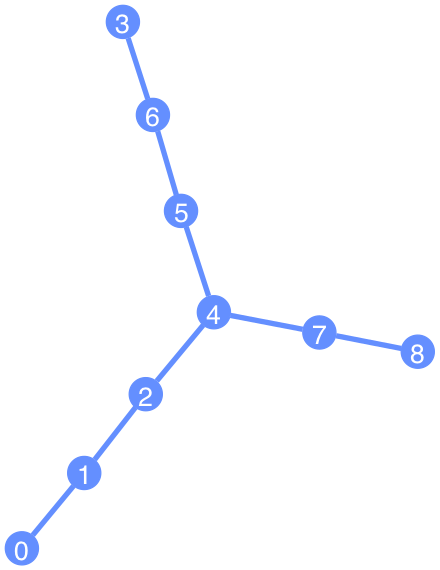} \label{fig:topology_9qY}} \\
            Topology ID: 9-T1 & Topology ID: 9-T2 & Topology ID: 9-Y\\
            &&\\
            {\includegraphics[height=0.25\columnwidth,valign=m]{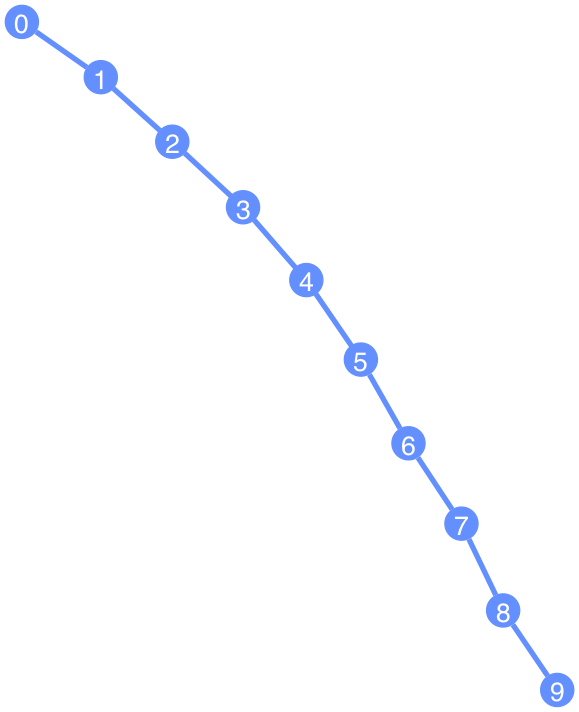} \label{fig:topology_10qL}} &{\includegraphics[height=0.25\columnwidth,valign=m]{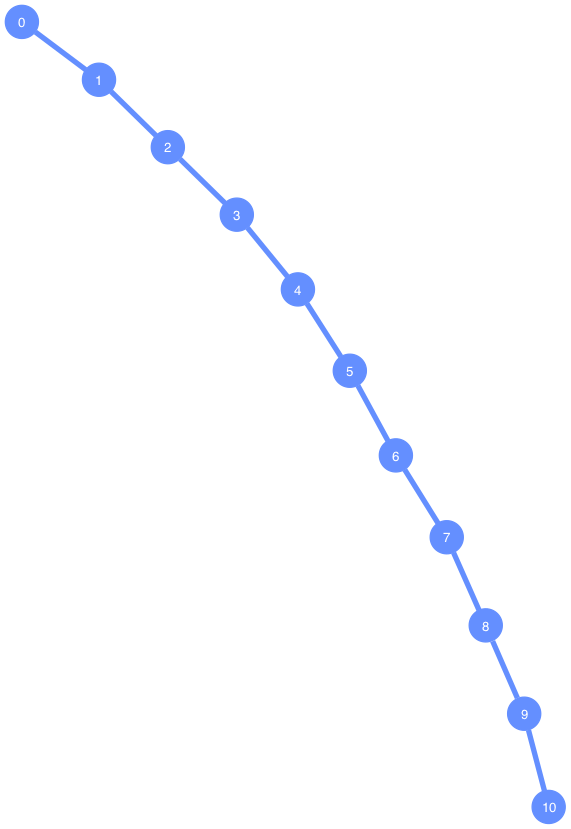} \label{fig:topology_11qL}} &
            {\includegraphics[width=0.25\columnwidth,valign=m]{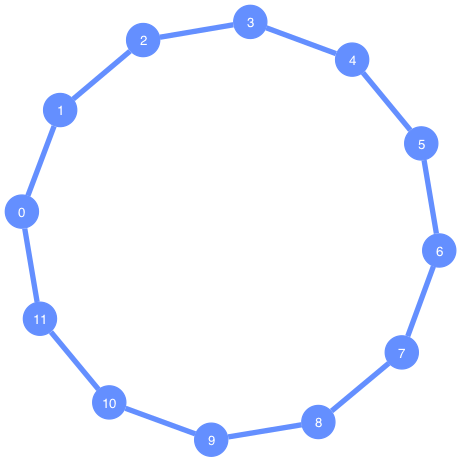}} \label{fig:topology_12qO} \\
            Topology ID: 10-L & Topology ID: 11-L& Topology ID: 12-O \\
        \end{tabular}
\end{table}

\begin{figure*}[h]
\centering
\includegraphics[width=0.7\linewidth]{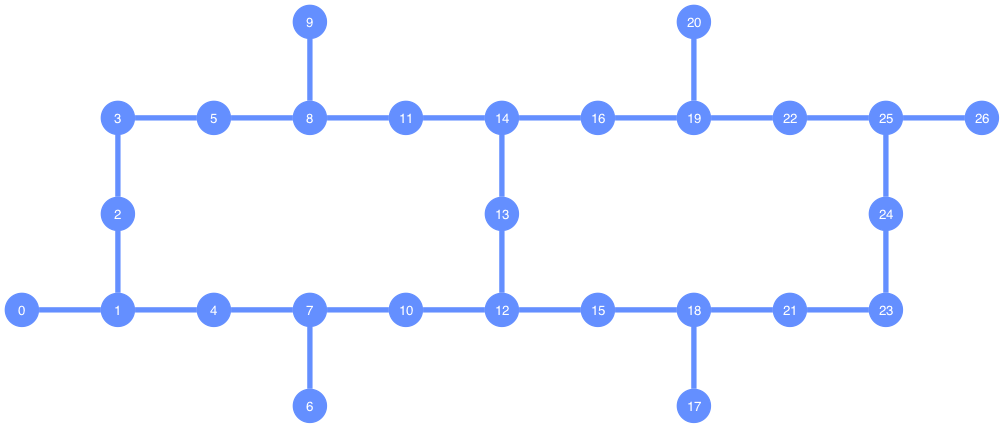}
\\ID: 27-HH. Coupling map: [[0, 1], [1, 0], [1, 2], [1, 4], [2, 1], [2, 3], [3, 2], [3, 5], [4, 1], [4, 7], [5, 3], [5, 8], [6, 7], [7, 4], [7, 6], [7, 10], [8, 5], [8, 9], [8, 11], [9, 8], [10, 7], [10, 12], [11, 8], [11, 14], [12, 10], [12, 13], [12, 15], [13, 12], [13, 14], [14, 11], [14, 13], [14, 16], [15, 12], [15, 18], [16, 14], [16, 19], [17, 18], [18, 15], [18, 17], [18, 21], [19, 16], [19, 20], [19, 22], [20, 19], [21, 18], [21, 23], [22, 19], [22, 25], [23, 21], [23, 24], [24, 23], [24, 25], [25, 22], [25, 24], [25, 26], [26, 25]]
\label{fig:topology_27qHH}
\end{figure*}

\begin{figure*}[tbp]
\centering
\includegraphics[width=0.4\linewidth]{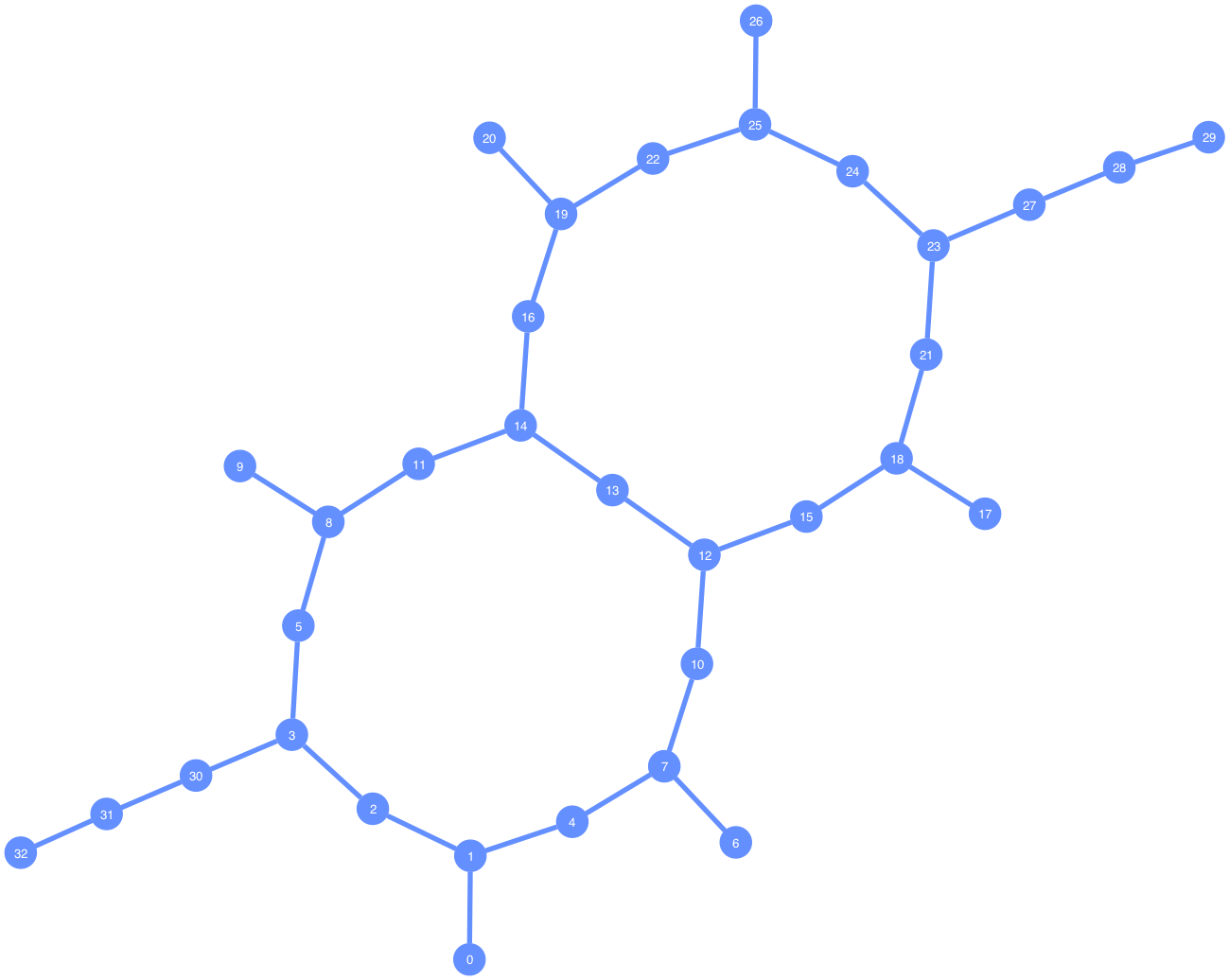}
\\ID: 33-HH. Coupling map: [[0, 1], [1, 0], [1, 2], [1, 4], [2, 1], [2, 3], [3, 2], [3, 5], [3, 30], [4, 1], [4, 7], [5, 3], [5, 8], [6, 7], [7, 4], [7, 6], [7, 10], [8, 5], [8, 9], [8, 11], [9, 8], [10, 7], [10, 12], [11, 8], [11, 14], [12, 10], [12, 13], [12, 15], [13, 12], [13, 14], [14, 11], [14, 13], [14, 16], [15, 12], [15, 18], [16, 14], [16, 19], [17, 18], [18, 15], [18, 17], [18, 21], [19, 16], [19, 20], [19, 22], [20, 19], [21, 18], [21, 23], [22, 19], [22, 25], [23, 21], [23, 24], [23, 27], [24, 23], [24, 25], [25, 22], [25, 24], [25, 26], [26, 25], [27, 23], [27, 28], [28, 27], [28, 29], [29, 28], [30, 3], [30, 31], [31, 30], [31, 32], [32, 31]]
\label{fig:topology_33qHH}
\end{figure*}

\begin{figure*}[tbp]
\centering
\includegraphics[width=0.4\linewidth]{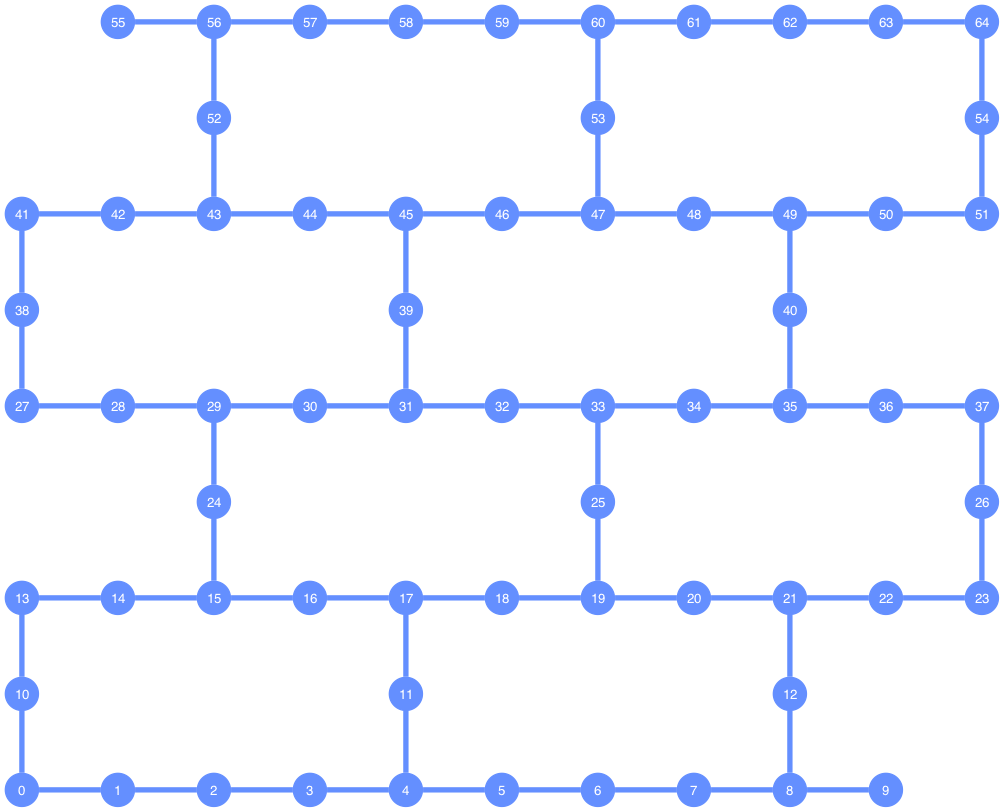}
\\ID: 65-HH. Coupling map: [[0, 1], [0, 10], [1, 0], [1, 2], [2, 1], [2, 3], [3, 2], [3, 4], [4, 3], [4, 5], [4, 11], [5, 4], [5, 6], [6, 5], [6, 7], [7, 6], [7, 8], [8, 7], [8, 9], [8, 12], [9, 8], [10, 0], [10, 13], [11, 4], [11, 17], [12, 8], [12, 21], [13, 10], [13, 14], [14, 13], [14, 15], [15, 14], [15, 16], [15, 24], [16, 15], [16, 17], [17, 11], [17, 16], [17, 18], [18, 17], [18, 19], [19, 18], [19, 20], [19, 25], [20, 19], [20, 21], [21, 12], [21, 20], [21, 22], [22, 21], [22, 23], [23, 22], [23, 26], [24, 15], [24, 29], [25, 19], [25, 33], [26, 23], [26, 37], [27, 28], [27, 38], [28, 27], [28, 29], [29, 24], [29, 28], [29, 30], [30, 29], [30, 31], [31, 30], [31, 32], [31, 39], [32, 31], [32, 33], [33, 25], [33, 32], [33, 34], [34, 33], [34, 35], [35, 34], [35, 36], [35, 40], [36, 35], [36, 37], [37, 26], [37, 36], [38, 27], [38, 41], [39, 31], [39, 45], [40, 35], [40, 49], [41, 38], [41, 42], [42, 41], [42, 43], [43, 42], [43, 44], [43, 52], [44, 43], [44, 45], [45, 39], [45, 44], [45, 46], [46, 45], [46, 47], [47, 46], [47, 48], [47, 53], [48, 47], [48, 49], [49, 40], [49, 48], [49, 50], [50, 49], [50, 51], [51, 50], [51, 54], [52, 43], [52, 56], [53, 47], [53, 60], [54, 51], [54, 64], [55, 56], [56, 52], [56, 55], [56, 57], [57, 56], [57, 58], [58, 57], [58, 59], [59, 58], [59, 60], [60, 53], [60, 59], [60, 61], [61, 60], [61, 62], [62, 61], [62, 63], [63, 62], [63, 64], [64, 54], [64, 63]]
\label{fig:topology_65qHH}
\end{figure*}

\begin{figure*}[tbp]
\centering
\includegraphics[width=0.4\linewidth]{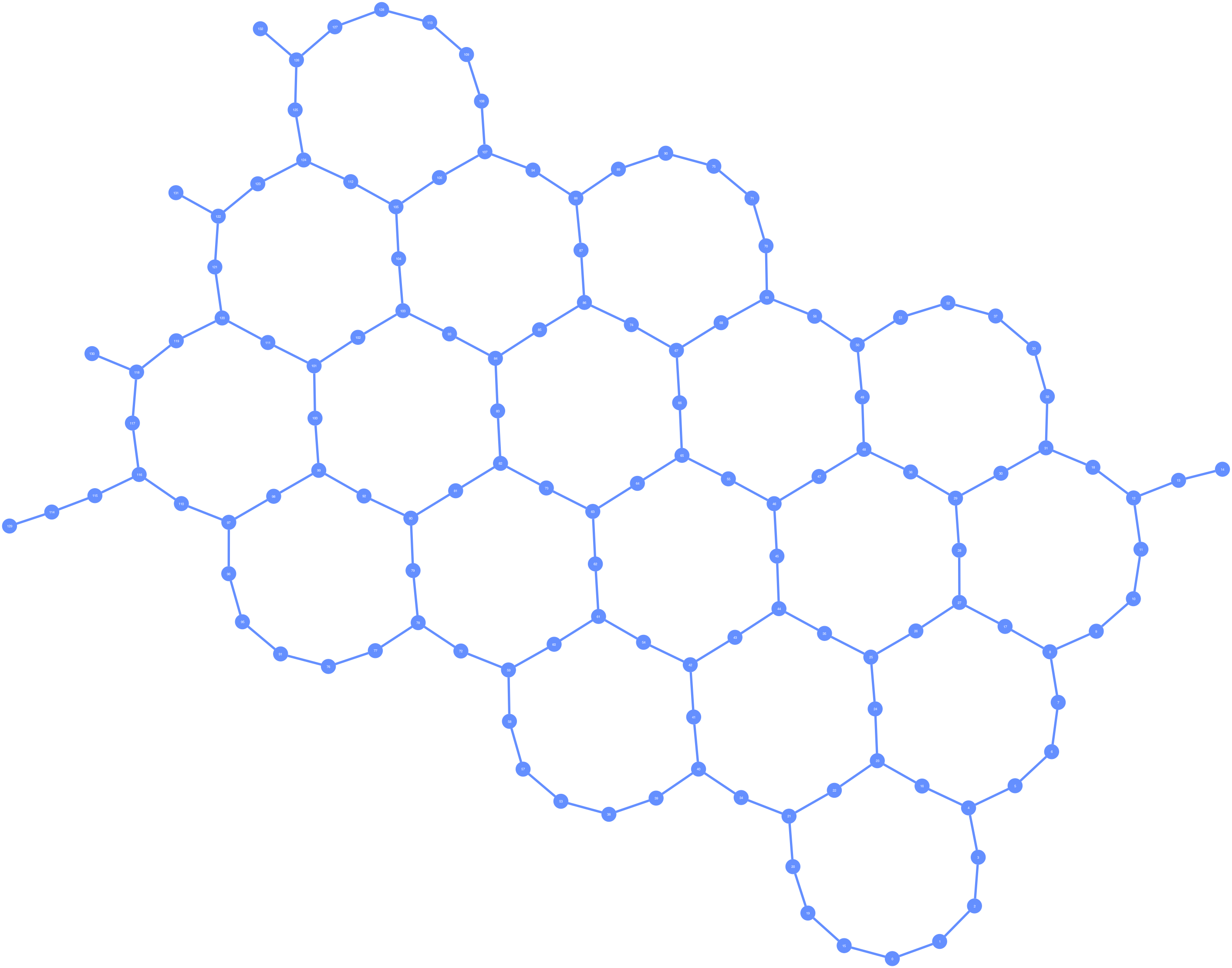}
\\Coupling map of \device{torino}: [[0,1], [0,15], [1,0], [1,2], [2,1], [2,3], [3,2], [3,4], [4,3], [4,5], [4,16], [5,4], [5,6], [6,5], [6,7], [7,6], [7,8], [8,7], [8,9], [8,17], [9,8], [9,10], [10,9], [10,11], [11,10], [11,12], [12,11], [12,13], [12,18], [13,12], [13,14], [14,13], [15,0], [15,19], [16,4], [16,23], [17,8], [17,27], [18,12], [18,31], [19,15], [19,20], [20,19], [20,21], [21,20], [21,22], [21,34], [22,21], [22,23], [23,16], [23,22], [23,24], [24,23], [24,25], [25,24], [25,26], [25,35], [26,25], [26,27], [27,17], [27,26], [27,28], [28,27], [28,29], [29,28], [29,30], [29,36], [30,29], [30,31], [31,18], [31,30], [31,32], [32,31], [32,33], [33,32], [33,37], [34,21], [34,40], [35,25], [35,44], [36,29], [36,48], [37,33], [37,52], [38,39], [38,53], [39,38], [39,40], [40,34], [40,39], [40,41], [41,40], [41,42], [42,41], [42,43], [42,54], [43,42], [43,44], [44,35], [44,43], [44,45], [45,44], [45,46], [46,45], [46,47], [46,55], [47,46], [47,48], [48,36], [48,47], [48,49], [49,48], [49,50], [50,49], [50,51], [50,56], [51,50], [51,52], [52,37], [52,51], [53,38], [53,57], [54,42], [54,61], [55,46], [55,65], [56,50], [56,69], [57,53], [57,58], [58,57], [58,59], [59,58], [59,60], [59,72], [60,59], [60,61], [61,54], [61,60], [61,62], [62,61], [62,63], [63,62], [63,64], [63,73], [64,63], [64,65], [65,55], [65,64], [65,66], [66,65], [66,67], [67,66], [67,68], [67,74], [68,67], [68,69], [69,56], [69,68], [69,70], [70,69], [70,71], [71,70], [71,75], [72,59], [72,78], [73,63], [73,82], [74,67], [74,86], [75,71], [75,90], [76,77], [76,91], [77,76], [77,78], [78,72], [78,77], [78,79], [79,78], [79,80], [80,79], [80,81], [80,92], [81,80], [81,82], [82,73], [82,81], [82,83], [83,82], [83,84], [84,83], [84,85], [84,93], [85,84], [85,86], [86,74], [86,85], [86,87], [87,86], [87,88], [88,87], [88,89], [88,94], [89,88], [89,90], [90,75], [90,89], [91,76], [91,95], [92,80], [92,99], [93,84], [93,103], [94,88], [94,107], [95,91], [95,96], [96,95], [96,97], [97,96], [97,98], [97,110], [98,97], [98,99], [99,92], [99,98], [99,100], [100,99], [100,101], [101,100], [101,102], [101,111], [102,101], [102,103], [103,93], [103,102], [103,104], [104,103], [104,105], [105,104], [105,106], [105,112], [106,105], [106,107], [107,94], [107,106], [107,108], [108,107], [108,109], [109,108], [109,113], [110,97], [110,116], [111,101], [111,120], [112,105], [112,124], [113,109], [113,128], [114,115], [114,129], [115,114], [115,116], [116,110], [116,115], [116,117], [117,116], [117,118], [118,117], [118,119], [118,130], [119,118], [119,120], [120,111], [120,119], [120,121], [121,120], [121,122], [122,121], [122,123], [122,131], [123,122], [123,124], [124,112], [124,123], [124,125], [125,124], [125,126], [126,125], [126,127], [126,132], [127,126], [127,128], [128,113], [128,127], [129,114], [130,118], [131,122], [132,126]]
\label{fig:topology_ibm_torino}
\end{figure*}

\end{document}